\documentclass[preprint,journal]{vgtc}            

\usepackage{amsmath}
\usepackage{microtype}
\usepackage{enumitem}
\usepackage{textcomp}
\usepackage{tikz}
\usetikzlibrary{arrows.meta}

\onlineid{0}

\vgtccategory{Research}

\vgtcpapertype{evaluation}

\definecolor{revisionblue}{RGB}{0,0,0}

\long\def\revise#1{{\color{revisionblue}#1}}

\title{Emotional Engagement in Narrative Medical Visualization: \\An Electrodermal Activity and Eye-Tracking Study}

\author{
  \authororcid{Beatrice Budich}{0009-0004-8741-8422}, \authororcid{Laura Garrison}{0000-0001-7134-2006}, \authororcid{Marc Vaudel}{0000-0003-1179-9578}, 
  \authororcid{Jone Trovik}{0000-0002-3808-6407}, \authororcid{Florian Heinrich}{0000-0002-8169-3157}, 
  \authororcid{Bernhard Preim}{0000-0001-9826-9478}, \\ and \authororcid{Monique Meuschke}{0000-0003-2183-6619}
}
\authorfooter{
 \item
  	Beatrice Budich is with the University of Magdeburg.
  	E-mail: beatrice.budich@ovgu.de
 \item
  	Laura Garrison is with the University of Bergen.
  	E-mail: Laura.Garrison@uib.no
    \item
  	Marc Vaudel is with the University of Bergen.
  	E-mail: Marc.Vaudel@uib.no
    \item
  	Jone Trovik is with the University of Bergen.
  	E-mail: Jone.Trovik@uib.no
    \item
  	Florian Heinrich is with the University of Magdeburg.
  	E-mail: florian.heinrich@ovgu.de
     \item
  	Bernhard Preim is with the University of Magdeburg.
  	E-mail: bernhard@isg.cs.uni-magdeburg.de
    \item
  	Monique Meuschke is with the University of Magdeburg.
  	E-mail: meuschke@isg.cs.uni-magdeburg.de
}

\teaser{
  \centering
  \includegraphics[width=\linewidth, alt={Visual Abstract: Prototyping, Experimental Procedure, Instruments, and Data Analysis.}]{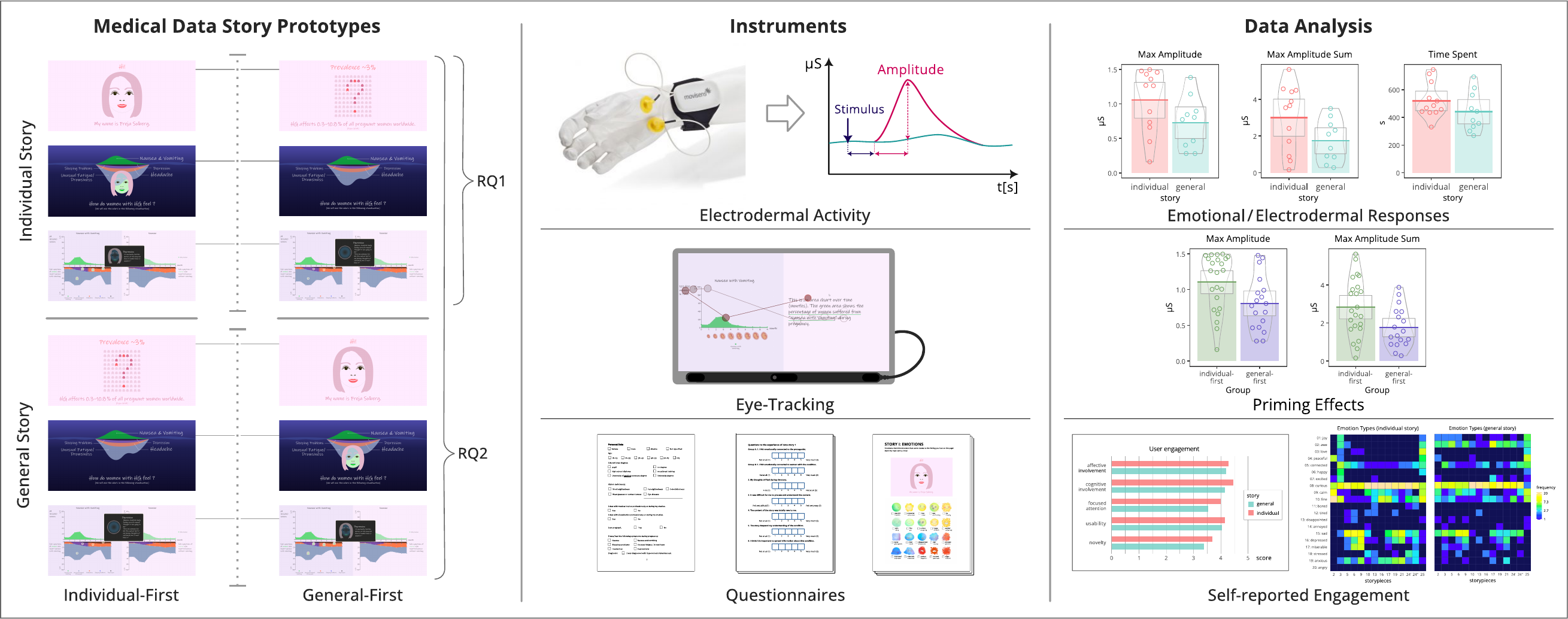}
  \caption{%
  	Overview of the study design, instruments, and analysis workflow. We presented two story prototypes (individual vs.\ general) in counterbalanced order, recorded EDA and eye-tracking data, and collected questionnaires for emotional response. We then computed derived metrics (peak amplitude, accumulated arousal, and time spent) to analyze engagement across story versions and segments.
  }
  \label{fig:VisualAbstract}
}

\abstract{%
Narrative visualization embeds data in visual stories to make medical information more relatable for non-experts. Despite the growing use of character elements in health communication, evidence on whether individual characters support affective responses remains inconclusive.
\revise{Physiological evidence independent of verbal self-report is especially scarce, although such measures should complement participants' self-reported experiences.}
We present a mixed-methods study using electrodermal activity (EDA), eye tracking, and questionnaires to compare two medical data stories: an individual, character-based version and a general, population-level version without an individual protagonist.
\revise{We examine how this framing influences physiological arousal, visual attention, viewing behavior, and self-reported affective response.}
\revise{Story-level EDA comparisons provide only limited support for stronger arousal in the character-based story. Stronger evidence comes from eye-tracking-based peak classification, where character illustrations were robustly associated with EDA peaks, and from questionnaire responses showing more negative empathy-related emotions for the individual story.}
\revise{The general story elicited more awe and joy, suggesting that individual and population-level framings may support different emotional qualities. We also observed a preliminary story-order effect: participants who first saw the individual story showed higher peak-based physiological arousal, although this effect cannot be fully disentangled from fatigue, novelty, or learning effects.}
By combining physiological, gaze-based, questionnaire, and lightweight qualitative evidence, our work advances time-resolved assessment of narrative medical visualization and highlights the need to interpret arousal, attention, curiosity, and self-reported emotions together.}

\keywords{Narrative Visualization, Electrodermal Activity, Eye Tracking, Data Analysis}

\graphicspath{{figs/}{figures/}{pictures/}{images/}{./}} 

\usepackage{tabu}                      
\usepackage{booktabs}                  
\usepackage{lipsum}                    
\usepackage{mwe}                       

\usepackage{mathptmx}                  

\begin{document}

\firstsection{Introduction}

\maketitle

Medical information is often complex, emotionally stirring, and hard to interpret. Narrative medical visualization embeds data in stories to make health information more accessible and support informed decision-making~\cite{meuschke2022narrative}. A common strategy is to use characters, alongside content, conflict, and structure, to foster emotional connection and prosocial responses such as empathy and support~\cite{mittenentzwei2023disease}. \revise{Because emotions can be described along valence (pleasant $\leftrightarrow$ unpleasant) and arousal (low $\leftrightarrow$ high), we distinguish throughout the paper between physiological arousal, visual attention, viewing time, cognitive-affective responses such as curiosity, and self-reported affect.}

Prior research identifies emotional engagement as a strong predictor of narrative impact~\cite{bilandzic2019narrative}, but the role of individual protagonists in narrative medical visualization remains inconclusive. Mittenentzwei et al.~\cite{mittenentzwei2023disease}, for instance, show that a patient character can increase engagement without affecting self-reported emotions. Few studies directly examine emotional engagement~\cite{airaldi2025best, mahyar2015towards}, and most rely on post-hoc self-reports~\cite{amini2018hooked, meuschke2022narrative, mittenentzwei2023disease, garreton2024attitudinal, boy2017showing}. \revise{Self-reports remain essential for capturing interpreted experiences and emotion labels, but they are retrospective and cannot show when arousal occurs during story viewing without interrupting the narrative~\cite{frey2020physiologically, samur2024getting}. EDA can enrich them through continuous, time-resolved arousal indicators independent of verbal reporting. By ``objective'', we mean physiologically measurable and self-report-independent, not direct or context-free access to emotion.}

\revise{We address these gaps with a data story about Hyperemesis Gravidarum (HG), a rare pregnancy condition marked by severe nausea and vomiting~\cite{fejzo2019nausea}. We compare a personalized version with a fictitious protagonist to a generalized version, examining physiological arousal, self-reported affect, and possible story-order effects. We focus on arousal because narrative medical visualizations may elicit heightened attention, concern, curiosity, or empathy. Since EDA cannot identify valence or specific emotions, we interpret it with self-reported emotions and eye-tracking data. Fig.~\ref{fig:VisualAbstract} summarizes this mixed-method workflow.} Our contributions are:

\begin{itemize}[noitemsep, topsep=0pt, parsep=0pt, partopsep=0pt,leftmargin=1.5em]
\item \revise{Two comparable HG data-story versions that manipulate character-based personalization (first-person, character-based) versus generalized population framing (third-person), while keeping the disease topic, narrative sequence, visualization structure, and interaction pattern closely matched.}
\item \revise{A mixed-methods study (N=26) combining EDA, eye tracking, questionnaires, and qualitative feedback to study physiological arousal, viewing behavior, and self-reported affective responses.}
\item \revise{A time-resolved analysis workflow that contrasts average and peak-sensitive physiological measures and links EDA peaks to story pieces and visual elements using eye-tracking data.}
\end{itemize}

\noindent \revise{Our results provide limited story-level EDA support for stronger arousal in the individual story after correction, but stronger convergent evidence that character illustrations are associated with EDA peaks and that the individual story elicits more negative empathy-related self-reports. The general story elicited more awe and joy, pointing to the complementary strengths of individual- and population-level framing. We also report a preliminary story-order effect that should be interpreted with caution, as it may reflect fatigue, novelty, learning, or cognitive effort.}

\section{Background}
\label{sec:back}
\revise{This section outlines the conceptual background for the study: emotional engagement, story-order effects, physiological arousal measures, and the medical context of HG. Technical details of EDA signals, preprocessing, and derived measures are consolidated in Sec.~\ref{sec:data-analysis}.}

\subsection{Foundations of Emotional Engagement}

\noindent\textbf{Emotional Engagement. }\label{sec:engagement}
Cavanagh et al.~\cite{cavanagh2019make} describe engagement as the “necessary first step for learning.” Engagement improves communication effectiveness~\cite{liem2020structure}, fosters positive emotions linked to satisfaction and continued use~\cite{amini2018hooked, seo2015users}, and supports enjoyment, persuasion, and identification~\cite{busselle2009measuring}. In healthcare, narratives promote empathy and reflection~\cite{liao2023narrative}, insight generation~\cite{boy2015can}, and learning motivation~\cite{obrien2008user, ozhan2020effects}, with sustained engagement leading to broader cognitive, emotional, and behavioral benefits~\cite{yardley2016understanding}.
Engagement is commonly used to assess the experience, effectiveness, and impact of data-driven stories~\cite{obrien2009development, mckenna2017visual}, but lacks a unified definition~\cite{mckenna2017visual}. Broadly, it spans affect, behavior, and cognition--the "ABCs of psychology"~\cite{lench2013functional}. Psychological Engagement Theory further defines it as emotional and cognitive connection to an activity aimed at meaningful outcomes~\cite{rachmad2022psychological}.

\revise{Here, \emph{emotional engagement} refers to viewers' affective involvement with a narrative, including emotional activation and subjective interpretation of content such as empathy, concern, curiosity, or resonance. Since it cannot be observed through a single modality, we distinguish it from our empirical indicators: EDA captures physiological arousal, eye tracking contextualizes viewed story elements, viewing time captures behavioral time-on-task, and questionnaires provide self-reported emotion labels and reflective interpretation. We interpret emotional engagement only when the indicators converge; otherwise, we refer directly to the measured construct.}


\noindent\textbf{Emotional Priming. }\label{sec:priming}
As we examine story-order effects, how story order influences physiological arousal, 
we briefly introduce emotional priming: the effect whereby a prior stimulus shapes perception of a later one. For example, negative facial expressions bias neutral faces negatively, while positive expressions bias them positively. These effects are strongest when prime and target are semantically related and share a modality, such as images~\cite{suslow2013neural}.


Emotional priming operates largely unconsciously and influences recall, judgment, and decision-making~\cite{si2024impact}. Emotional cues attract more attention than neutral stimuli and can enhance attention and memory~\cite{pessoa2005extent, pessoa2002attentional}. However, effects depend on valence: positive stimuli often improve memory, whereas negative emotions may lower recognition thresholds and distort recall~\cite{windmann2001electrophysiological}.
Emotional priming has also been examined in storytelling. Rao et al.~\cite{rao2019effects} showed that participants primed with emotional expressions, either by viewing or enacting them, produced richer stories than in a no-prime condition, with greater idea diversity and linguistic variety. This suggests that emotional primes at the beginning of a story can shape and intensify narrative expression.

\subsection{Arousal Measures in Narrative Evaluation}\label{sec:basics-eda}
\revise{EDA has been used in psychophysiology, HCI, VR, learning, and narrative research as a non-invasive indicator of physiological arousal~\cite{boucsein2012electrodermal, cowley2016short, frey2020physiologically, richardson2018measuring}. For narrative visualization, its main value is that it can be recorded continuously during story viewing and help locate moments of heightened arousal without interrupting the narrative. However, EDA does not identify emotional valence or emotion type on its own. Elevated arousal may reflect affective response, attention, curiosity, cognitive effort, stress, surprise, or personal relevance, and may be shaped by prior health experience and broader biopsychosocial factors. We therefore use EDA as one component of a mixed-methods analysis and interpret it together with eye tracking, self-reported emotions, and qualitative feedback.}

\subsection{Hyperemesis Gravidarum (HG)}\label{sec:hg}
HG is a rare pregnancy condition marked by severe, persistent nausea and vomiting, with reported prevalence ranging from 0.3–10.8\% across countries and ethnic groups~\cite{fejzo2019nausea}. Affected women often require hospitalization for hydration, nutrition, and medication, and many need psychological support due to substantial mental strain. Although potentially life-threatening because of metabolic and electrolyte imbalances~\cite{bailit2005hyperemesis}, HG is frequently misdiagnosed as normal pregnancy nausea with psychological causes~\cite{havnen2019women}. Limited awareness of its biological and genetic factors and underuse of treatment options continue to endanger both mothers and their unborn children~\cite{fejzo2024hyperemesis}.

\section{Related Work}
\label{sec:relwork}
\revise{This section situates our work within prior research on narrative medical visualization, character design, and physiological measures for evaluating narrative experiences.}

\subsection{Narrative Medical Visualization}\label{sec:nvm}
Segel and Heer~\cite{segel2010narrative} established narrative visualization through an analysis of 58 data journalism examples and a framework of genres and narrative strategies, but did not evaluate engagement or comprehension. In healthcare, narrative visualization adds challenges because complex medical data must be communicated to non-experts with clarity and sensitivity. While infographics and dashboards support comprehension~\cite{bhat2023infographics}, narrative visualization may further enhance medical communication.

Meuschke et al.~\cite{meuschke2022narrative} introduced narrative medical visualization for communicating complex medical data to non-experts through storytelling and proposed a seven-stage tension-arc template for disease stories. In a questionnaire-based study (N=90), interactive medical data stories improved comprehension, retention, and affective involvement compared to a blog format, especially after introducing a patient. However, the study could not isolate effective narrative elements or analyze engagement over time, despite suggesting that stronger patient focus could enhance emotional engagement.

Evaluation methods in narrative visualization vary widely, from surveys~\cite{mittenentzwei2023disease} to interviews~\cite{lan2022negative} and detailed emotion scales~\cite{nowak2018micro}. Yet most studies neglect emotional changes over time and rarely distinguish arousal (intensity) from valence (pleasantness). \vspace{0.1cm}

\noindent
\textbf{Character Design.}
Research on data-driven storytelling and narrative visualization highlights the use of characters to structure narratives, guide interpretation, and humanize data, fostering empathy and accessibility~\cite{campbell2008hero,giovannelli2009sympathy}. Anthropomorphism can increase relatability and emotional impact, and boost empathy and engagement~\cite{lan2023affective,boy2017showing}. Storytelling techniques guide attention in exploratory visualizations~\cite{boy2015can}, while game-design research shows that expressive characters enhance immersion and narrative progression~\cite{cavazza2002interacting,mariani2019character,sheldon2022character}. Psychological work links these effects to emotional contagion and “social syncing”~\cite{isbister2022better}. In visualization, Dasu et al.~\cite{dasu2023character} show that characters structure data-driven stories and improve accessibility, though many approaches still require manual design and lack expressive or automated generation.

In medical storytelling, characters often represent patients and contextualize health data~\cite{mittenentzwei2023disease,neeley2020linking}. Designers may derive anonymized yet realistic patient representations from epidemiological datasets such as SHIP~\cite{volzke2022cohort,budich2023reflections}, supporting individual storytelling while addressing privacy constraints. However, convincing character creation still requires manual effort, interdisciplinary expertise, and iterative workflows~\cite{budich2023reflections,mittenentzwei2024ai}. Generative AI, including Midjourney~\cite{midjourney2024} and Stable Diffusion~\cite{stabilityai2024}, 
enables semi-automated prompt-based workflows~\cite{liu2023pre}, but prior work reports limited visual consistency, user evaluation, emotional expression, and animation quality~\cite{budich2023reflections,mittenentzwei2024ai}. Hybrid approaches combining 3D modeling, animation, and real patient input aim to strengthen identification~\cite{budich2025narrative}.

Empirical research on characters in medical data storytelling remains limited. Mittenentzwei et al.~\cite{mittenentzwei2023disease} show that narratives with a protagonist, especially patient-centered ones, increase engagement but do not influence behavioral change. Their study uses a small sample and post-hoc self-reports, without capturing emotional dynamics.
%

Overall, prior work shows growing interest in character design for medical data stories, but lacks detailed analysis of emotional dynamics and relies mainly on self-reports.

\subsection{Measuring Physiological Arousal with EDA}
\label{sec:measuring-eda}
Narrative visualization research largely relies on subjective measures, while 
EDA offers a non-invasive, continuous indicator of physiological arousal, commonly used in stress and engagement research~\cite{patel2024electrodermal,cowley2016short}.

\revise{Findings on relations between self-reports and EDA are mixed. Lang et al.~\cite{lang1993looking} reported agreement between self-reported arousal/valence and EDA responses to emotional images, whereas Richardson et al.~\cite{richardson2018measuring} found higher physiological arousal for audiobooks despite higher self-reported engagement for videos, suggesting divergence between emotional and cognitive engagement. Similarly, Stuhldreher and Brouwer~\cite{stuldreher2025monitoring} found no correlation between self-reported attentional engagement and EDA during a lecture. These studies show that EDA should not replace self-report: it helps locate arousal over time, while self-report remains necessary for emotion type and valence.}

Several studies highlight EDA’s potential in adaptive and immersive contexts. Frey et al.~\cite{frey2020physiologically} used EDA and eye tracking for real-time story adaptation, achieving 92.9\% arousal classification accuracy. Rühlemann~\cite{ruehlemann2022emotional} found that audience arousal mirrored a storyteller’s arousal intensity, suggesting empathy through mirroring. In VR learning, Liberman and Dubovi~\cite{liberman2022effect} found that spoken narration increased EDA peaks and cognitive load without improving learning, while Liberman et al.~\cite{liberman2023impact} reported that high-immersion VR raised arousal and supported declarative knowledge, whereas low immersion benefited procedural learning.
Ma et al.~\cite{ma2023understanding} used EDA, eye tracking, and facial expression analysis to study engagement in an online statistics course. Physiological data captured arousal and attention, but no significant correlations emerged between EDA peaks and self-reported emotions. Active elements and animations increased perceived engagement, highlighting the complexity of linking physiological signals to subjective experience.

\revise{Overall, prior work highlights the potential and challenges of EDA. It provides continuous physiological information about arousal that may diverge from self-reports because arousal can reflect emotion, attention, learning, curiosity, stress, or cognitive effort. Crucially, no studies have applied EDA to narrative visualization or linked peak-sensitive physiological measures to story pieces and viewed elements over time.}

\section{Study Design}\label{sec:experiment} 

\begin{figure*}[t]
    \centering
    \includegraphics[width=0.95\linewidth, alt={}]{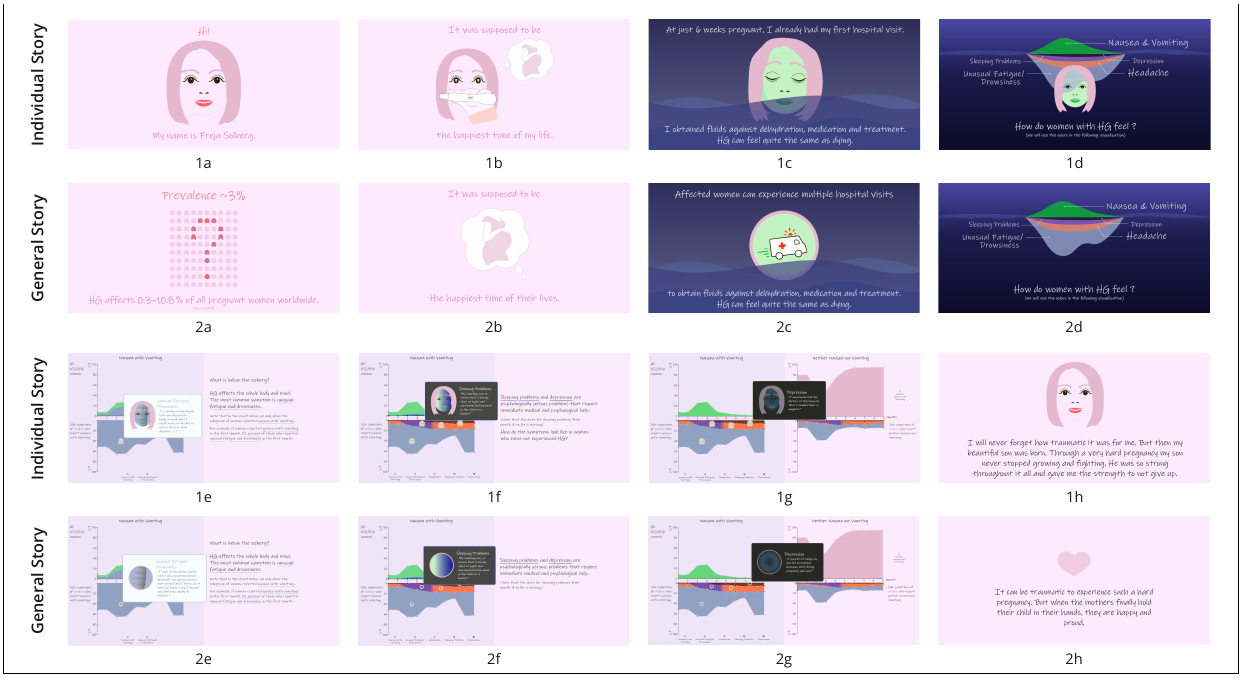}
    \caption{\revise{Selected story pieces from the individual (1a--1h) and the general story (2a--2h). Full story versions are provided in the supplemental material. }}
    \label{fig:prototypes}
    \vspace{-10px}
\end{figure*}

\revise{This study examines whether a character-based data story is associated with different physiological arousal, viewing behavior, and self-reported affective responses than a general-perspective version. It also examines whether story order is associated with subsequent responses.}

\subsection{Key Research Questions \& Hypotheses}

Based on the gaps we identified in previous research, our study addresses the following research questions: \vspace{0.1cm}

\noindent
\textbf{RQ1:} \textit{\revise{Does a fictitious, individual character change viewers' physiological arousal and self-reported affective responses to a medical data story compared to a general story without a human protagonist?}}

\noindent
\textbf{RQ2:} \textit{\revise{Does story order influence physiological arousal when participants view the individual and general versions sequentially?}}\vspace{0.1cm}

\noindent We formulate the following hypotheses to address our RQs: 
\begin{itemize}[noitemsep, topsep=0pt]
  \item [\textbf{H1: }] Including a fictitious, individual character who embodies a patient's lived experience will elicit increased \revise{physiological} arousal from the audience.  
  \item [\textbf{H2:  }] The individual story elicits stronger negative empathic emotions (sad, depressed, miserable, anxious) than the general story.
  \item [\textbf{H3:  }] Participants who read the individual story first, followed by the general story, will exhibit higher \revise{physiological} arousal than participants who read the stories in reverse order.
\end{itemize}

\noindent
We employed a mixed study design with one within-subjects factor (story version: individual vs. general) and one between-subjects factor (viewing order: \revise{individual-first vs. general-first}), with counterbalanced stimulus order. \revise{Participants were alternately assigned to the two order groups during recruitment to keep group sizes balanced:}
\begin{itemize}[noitemsep, topsep=0pt, parsep=0pt, partopsep=0pt,leftmargin=1.5em]
    \item \textbf{\revise{\textit{Individual-first group:}}} (1) individual story, then (2) general story; 
    \item \textbf{\revise{\textit{General-first group:}}} (1) general story, then (2) individual  story.
\end{itemize}

\noindent 
The \textit{independent variables} are defined as the presence of an individual protagonist (RQ1) and story sequencing (RQ2). The \textit{dependent variables} are \revise{EDA-derived physiological arousal measures, viewing time, eye-tracking-based peak labels, and self-reported affective responses.}

\subsection{Stimuli Design}\label{subsec:story}
We first describe the story concept and similarities between versions. Both are web-based slideshows composed of slides, referred to as \emph{story pieces}~\cite{lee2015more}. We then highlight the differences between the two designs.

\subsubsection{Story Concept and Similarities in the Stimuli}
\label{subsubsec:stimuli}
The narrative intent of both stories is identical: to communicate HG as a serious biologically based pregnancy condition, distinguish it from ``normal'' nausea, and promote earlier detection, medical support, everyday care, and empathy for affected women~\cite{fejzo2019nausea}.
Accordingly, the target audience includes partners, relatives, friends, and a broader audience. Both versions are based on cohort data from the Norwegian Mother, Father and Child Cohort Study (MoBa) (102,810 women)~\cite{brandlistuen2025cohort, moba}. All charts show maternal symptom prevalence by pregnancy week (based on responses to questionnaires 1 and 3, MoBa). For details on MoBa data and preparation, see App.~\ref{app:Moba} in the supplemental material. 

\noindent\textit{Narrative Structure:} \revise{Both versions were designed to be comparable rather than perceptually identical. They follow the same five-phase narrative arc, use the same web-based slideshow interaction, contain the same core information and data visualizations, and introduce the same medical explanations and population-level MoBa data. However, they differ in visual and rhetorical framing: the individual story uses a named protagonist, first-person narration, facial expressions, and character-linked cues, whereas the general story uses third-person narration, icon-based imagery, population-level framing, and quotes from multiple women. We therefore interpret condition differences as effects of the broader character-based framing rather than of a single isolated visual variable.} The only systematic difference lies in how the narrative is visually personified, described in Sec.~\ref{subsubsec:visdiff}.

\vspace{0.1cm}
\noindent\textbf{Story Title:} \textit{"Surviving Hyperemesis Gravidarum"}

\begin{enumerate}[noitemsep, topsep=0pt, parsep=0pt, partopsep=0pt,leftmargin=1.5em]
    \item  \textbf{Introduction:} A positive pregnancy introduction sets the context.   
    \item \textbf{Escalation:} The conflict unfolds as moderate nausea escalates into a “nightmare”, highlighting rapid onset and need for treatment.
    \item \textbf{Explanation:} As education is central to the story's aims, we explain the biological and genetic aspects of HG with animated illustrations.
    \item \textbf{Data-driven Insights:} The MoBa dataset is introduced through an iceberg metaphor, illustrating that prominent symptoms like nausea and vomiting are visible above the waterline, while less visible symptoms lie beneath, e.g., fatigue, headache, sleeping problems, and depression. The chart is built incrementally across slides, including interactive tooltips that reveal symptom-specific quotes. Finally, two comparative visualizations contrast women with nausea and vomiting against those with only nausea or none of these symptoms, highlighting the greater severity of HG.
    \item \textbf{Resolution:} The story ends with a hopeful tone and a call to action.
\end{enumerate}

\subsubsection{Visualization Design}
\label{subsubsec:visdesign}
\revise{The central visualization depicts symptom prevalence across pregnancy weeks (Fig.~\ref{fig:prototypes}.1f and 2f). The x-axis encodes gestational week and the y-axis the proportion of MoBa participants reporting each symptom. Symptoms are introduced incrementally to reduce visual complexity and to support narrative pacing. The iceberg metaphor contrasts visible symptoms, such as nausea and vomiting, with less visible symptoms below the waterline, including fatigue, headache, sleeping problems, and depression. Tooltips link prevalence curves to symptom-specific quotes. Later story pieces compare women with HG, nausea and vomiting, and no nausea or vomiting, highlighting differences in symptom burden (Fig.~\ref{fig:prototypes}.1g and 2g). Larger exemplary images are provided in App.~\ref{app:story-screenshots} (Fig.~\ref{fig:results-story-level_1} and Fig.~\ref{fig:results-story-level_2}) in the supplemental material.}

\revise{We used the iceberg metaphor to show both publicly visible HG symptoms and less visible physical and psychological consequences, introducing the data progressively while emphasizing the often-underestimated disease burden. The visualization structure, data, axes, interaction pattern, and sequence of visual reveals were held constant across both stories. Because EDA peaks were later classified by viewed story elements, the stimuli were labeled with categories such as character illustration, data visualization, text, icons, and tooltip areas.}


\subsubsection{Differences between Stimuli}
\label{subsubsec:visdiff}

The two versions shared the same disease information and visualization logic, but differed in narrative and visual framing.

\noindent
\textbf{\textit{Individual Story (character-driven visuals):}} This story is anchored by a fictional patient character, told from the perspective of an individual protagonist. The character was inspired by a blog post on an Australian HG support website~\cite{emily}, which provided insights into the disease process and its emotional impact. 
The protagonist is visually present throughout the story, often occupying most of the slide to foreground emotion and embodied experience. The texts were closely aligned with the blog post and told in the first-person perspective. 

The protagonist's presence begins with a direct character introduction as Freja Solberg (Fig.~\ref{fig:prototypes}.1a), and continues through the early pregnancy framing (Fig.~\ref{fig:prototypes}.1b). As the narrative escalates, her facial depiction conveys symptom progression and distress, e.g., the hospitalization scene (Fig.~\ref{fig:prototypes}.1c). The character is also integrated into the narrative framing: the iceberg metaphor is combined with the protagonist’s presence (Fig.~\ref{fig:prototypes}.1d). In the \emph{Data-Driven Insights} phase, both versions use the same chart structure, showing aggregated MoBa data, while the individual story adds character-linked cues such as portrait-style icons and character-associated tooltips (Fig.~\ref{fig:prototypes}.1e--g). The closing returns to the protagonist’s personal outcome and emotional tone (Fig.~\ref{fig:prototypes}.1h).

\noindent
\textbf{\textit{General Story (icon-based, non-personal visuals):}} 
In contrast, the general version uses a third-person narration and generic, icon-based visuals, shifting the visual “protagonist” from an individual to the disease and the affected population. This appears in the prevalence depiction using a population-style icon (Fig.~\ref{fig:prototypes}.2a) and continues in the pregnancy framing without a character portrait (Fig.~\ref{fig:prototypes}.2b). Key events are communicated via \revise{non-personalized} pictograms, e.g., an ambulance icon (Fig.~\ref{fig:prototypes}.2c), and the iceberg is shown without a human figure (Fig.~\ref{fig:prototypes}.2d). In the \emph{Data-Driven Insights} phase, the story uses general icons and non-personal tooltips with quotes from multiple women (Fig.~\ref{fig:prototypes}.2e-g), ending with \revise{a non-personalized symbolic conclusion instead of a character-focused one (Fig.~\ref{fig:prototypes}.2h). Here, non-personalized means that the slide does not show a named protagonist, facial expression, or individual outcome, rather than that it lacks emotional impact.}

\subsection{Pilot Study}
We conducted a \revise{formative} qualitative pilot study as part of a human-centered design process to refine the prototypes and plan the final evaluation. \revise{The pilot was not a formal test of emotional engagement.} First, we evaluated the individual story with two participants: an HG researcher and a former patient. After refining and developing the general story, we evaluated both versions with four additional participants, including a former patient, a visualization expert, and \revise{two lay participants without professional expertise in HG, medicine, or visualization.}

Participants were briefed on the study procedure before viewing the stories on a laptop and asked to express immediate thoughts and feelings using a think-aloud protocol. A short interview was then conducted to capture overall experience and usability or comprehension issues. \revise{Former HG patients and the HG researcher mainly commented on medical accuracy, experiential plausibility, and whether the symptoms and emotional tone were recognizable from lived or professional experience. The visualization expert and lay participants focused on comprehensibility, pacing, usability, and overall reception.} The first two sessions were documented in writing, and the remaining four were audio-recorded.

\revise{During the pilot, ``engagement'' was not measured as a formal outcome. We used the term only for spontaneous indications of interest, emotional connection, attention, confusion, or disengagement in think-aloud comments and post-session feedback.  Participants reported interest in both story versions, especially quotes, data visualizations, and animations, while several described a stronger personal connection to the individual story. These observations informed design refinements and are not used as confirmatory evidence for the main study results.}

\subsection{Study Recruitment \& Procedure}\label{subsec:procedures}
We recruited participants through our personal networks and through  snowball sampling~\cite{goodman1961snowball}. 
Apart from basic English proficiency, no specific requirements were set. Key terms were explained in advance. 
Mothers were actively recruited to incorporate maternal perspectives.

\noindent\textbf{Participant demographics. } 
In total, 26 participants took part in the study. Three were excluded during data collection due to lost EDA sensor connection and replaced to maintain balanced groups. The final sample included 15 female and 11 male participants, mostly aged 26--35 (n=15). Others were 18--25 (n=4), 36--45 (n=6), or 46--55 (n=1). Five women were mothers, including one former HG patient, and one participant was pregnant.
Most participants held a university degree (n=21). Others had a university of applied sciences degree (n=1), high school diploma (n=3), or secondary school diploma (n=1).
Participants worked in industry (n=11), research (n=11), or were students (n=4). 
Six had experience in medicine and visualization, and two in visualization only. 
Among the six participants with pregnancy experience, three reported nausea, and one reported nausea with vomiting. Additional symptoms included sleep problems (n=2), fatigue (n=1), and headaches (n=1). The former HG patient experienced persistent nausea with vomiting and received medication.
For MoBa data availability and study ethics declarations, see App.~\ref{app:Moba} in the supplemental material. 

\noindent\textbf{Experimental setup. }  
We conducted the experiment in a climate-controlled lab with the same setup across sessions, alternating participants between the \revise{individual-first} and \revise{general-first} groups. Participants were offered refreshments and completed the study in 25--45 minutes.
%
\revise{T0--T8 denote sequential procedure stages, with activities and collected data listed in Tab.~\ref{tab:procedure}. Participants completed questionnaires at predefined points during the session, followed by an open, voluntary discussion and informal feedback.}

\begin{table}[t]
\caption{\revise{Study procedure and timing of data collection.}}
\label{tab:procedure}
\centering
\scriptsize
\begin{tabular}{p{0.08\columnwidth}p{0.40\columnwidth}p{0.40\columnwidth}}
\toprule
\textbf{Stage} & \textbf{Activity} & \textbf{Data collected} \\
\midrule
T0 & Briefing and informed consent & Consent \\
T1 & Demographic questionnaire & Demographics and background \\
T2 & EDA sensor setup & Physiological sensor setup \\
T3 & Eye-tracker calibration & Calibration data \\
T4 & Story 1 & EDA and eye tracking \\
T5 & Post-story questionnaire 1 & Emotion-selection and engagement questionnaires \\
T6 & Story 2 & EDA and eye tracking \\
T7 & Post-story questionnaire 2 & Emotion-selection and story-preference question \\
T8 & Open discussion & Optional informal feedback \\
\bottomrule
\end{tabular}
\end{table}

\section{Data Collection \& Analysis Workflow}
\label{sec:data-analysis}
\revise{This section summarizes the workflow used for hypothesis testing. Implementation details on EDA preprocessing, synchronization, labeling, and peak classification are provided in  App.~\ref{app:data-analysis-details}} in the supplemental material. Our workflow consisted of: (1) collecting and (2) preprocessing EDA and eye-tracking data, \revise{(3) synchronizing story navigation with EDA data  and labeling it by story version and story piece, (4) computing EDA-derived physiological arousal metrics, (5) using eye tracking to attribute EDA peaks to viewed story elements,} (6) analyzing questionnaires and lightweight qualitative feedback, and (7) statistical testing for H1--H3.

\subsection{Data Collection}
\revise{This section summarizes the technical EDA information. We used EDA as the primary continuous measure of physiological arousal. EDA reflects sympathetic changes in skin conductance and includes a slow tonic component, skin conductance level (SCL), and a faster phasic component, skin conductance response (SCR). SCR amplitude indicates arousal intensity, but not emotional valence or whether arousal reflects emotion or cognitive effort. We therefore analyze SCR amplitude alongside self-reported emotions and eye-tracking context.
}
\begin{figure}[t]              
    \centering
    \includegraphics[width=0.9\columnwidth]{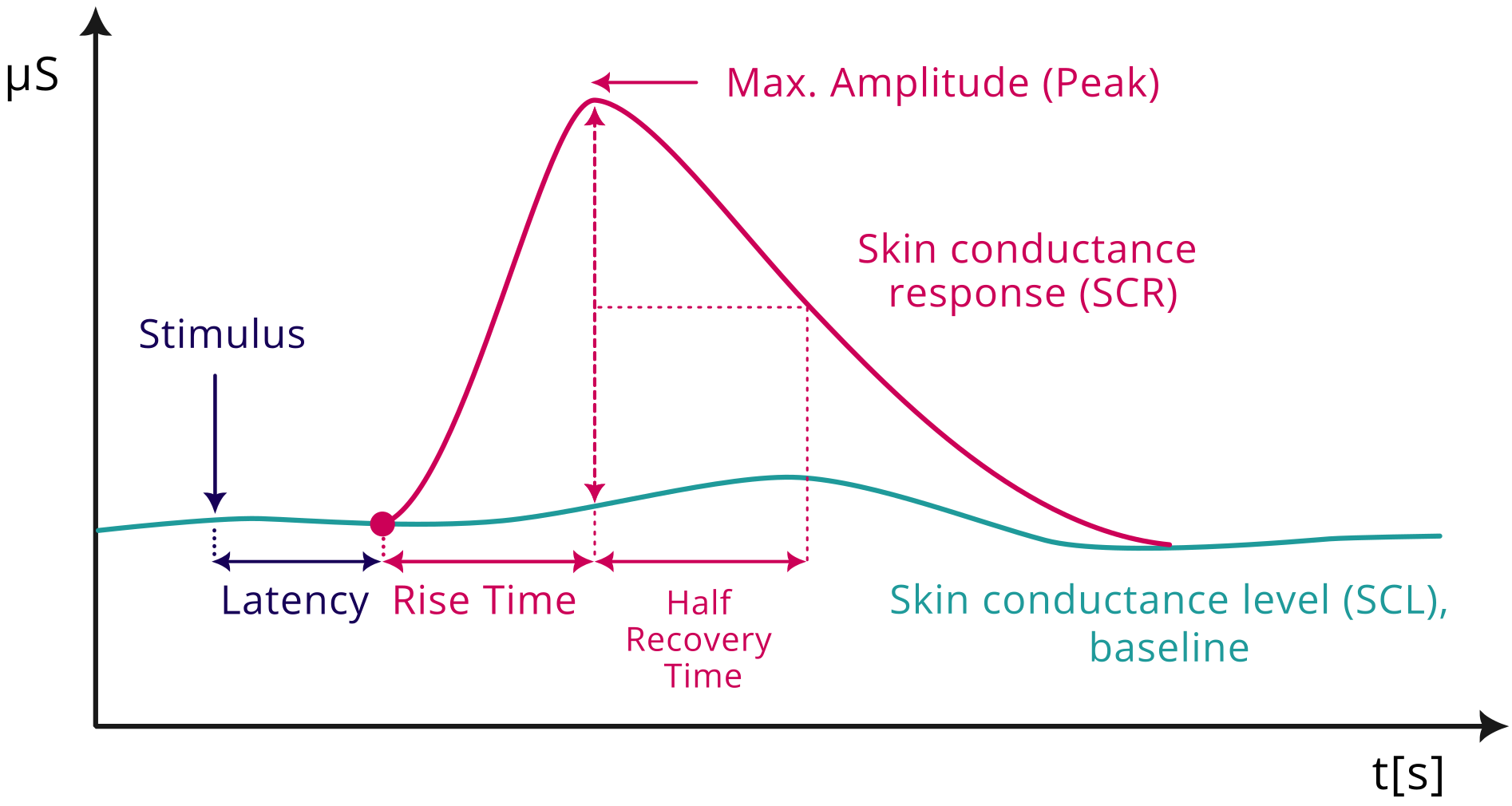}
    \caption{\revise{Schematic EDA signal showing tonic skin conductance level (SCL), phasic skin conductance response (SCR), latency, rise time, maximum amplitude, and recovery.} [\textit{created by the authors}] 
    }
    \label{fig:eda-signal}
       \vspace{-0.8em}
\end{figure}
\revise{
Fig.~\ref{fig:eda-signal} summarizes the EDA components used in our analysis: tonic SCL for data-quality checks and phasic SCRs for peak-sensitive arousal measures and latency-corrected peak labels.}

\begin{figure}[t]              
    \centering
    \includegraphics[width=0.5\columnwidth, alt={}]{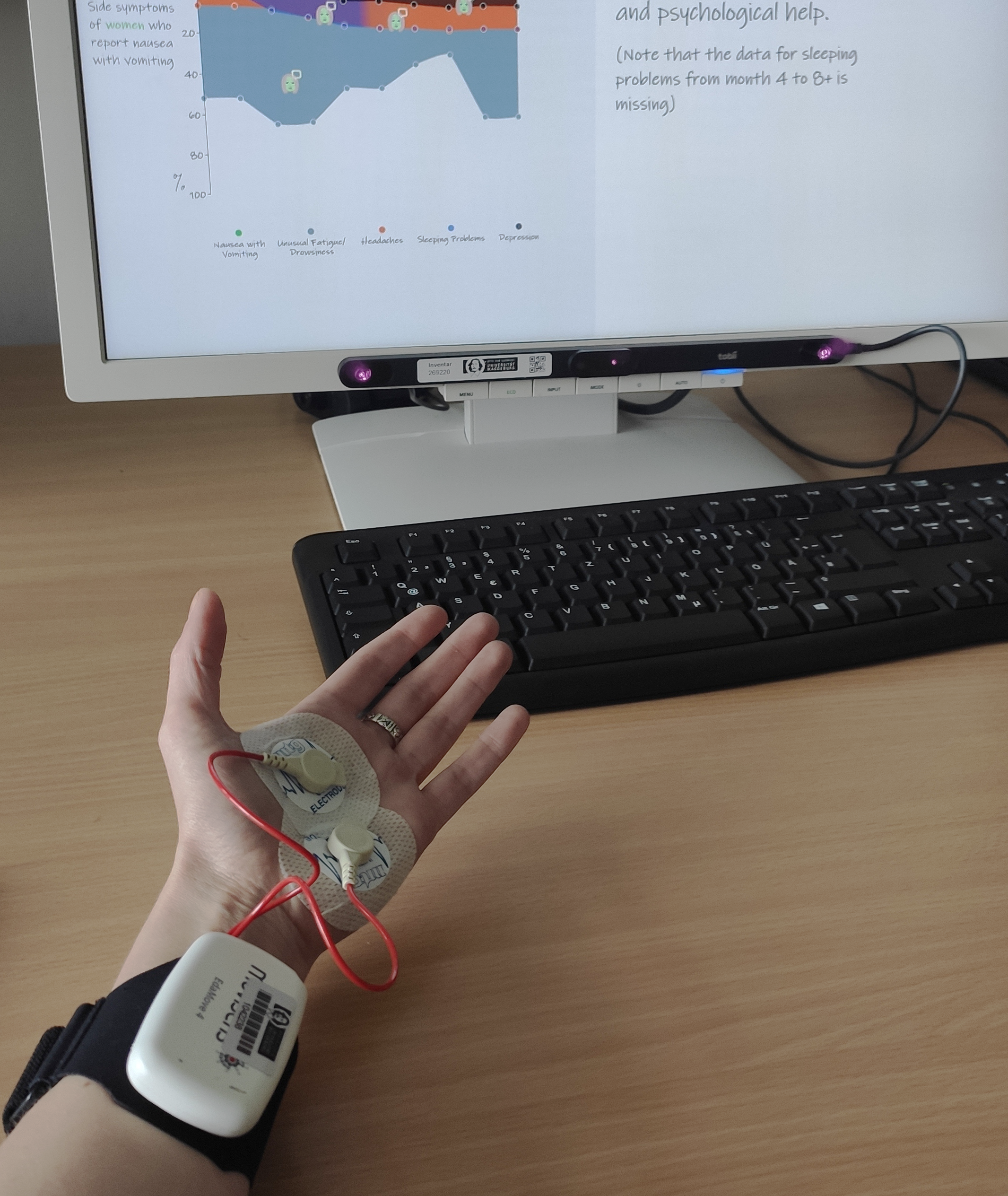}
    \caption{Experimental setup and biophysical data collection instruments.}
    \label{fig:setup}
    \vspace{-0.5em}
\end{figure}

\noindent\textbf{EDA.} We collected EDA data using the wearable EdaMove 4 sensor~\cite{EdaMove4} by movisens GmbH~\cite{movisens}, which records EDA, skin temperature, physical activity, position, and acceleration. Its wearable design allows natural movement, so participants were not required to keep their hands still during the experiment. The sensor uses the Exosomatic Measurement Method~\cite{society2012publication} and applies a Direct Current voltage of 0.5V to the skin to measure conductance between two electrodes placed on the palms of the hands (Fig.~\ref{fig:setup}). The sensor measures skin conductance in microsiemens (\textmu S). Automated skin conductance response detection achieves 92\% sensitivity, although peaks $>$2\textmu S may be underestimated by 5--10\% due to 14-bit quantization and 32Hz sampling.


\noindent\textbf{Eye Tracking.} \revise{We used eye tracking to synchronize scroll-based navigation, identify visible story pieces, and attribute EDA peaks to viewed elements.} The Tobii Pro Spark~\cite{tobii-spark}, mounted below the monitor (Fig.~\ref{fig:setup}), recorded gaze data at 60Hz with synchronized screen recordings collected in Tobii Pro Lab~\cite{tobii-lab}.


\noindent\textbf{Participant Data.} We included all 26 participants in the questionnaire analyses. For biophysical analyses, we excluded three participants with $>$10\% missing EDA data and the former HG patient, whose extreme amplitudes would have disproportionately influenced the small-sample comparisons. \revise{We return to this case in the limitations as an example of how lived experience may shape physiological responses to health narratives.} This resulted in 12 participants in the individual-first group and 10 in the general-first group.

\subsection{EDA Preprocessing \& Data Quality Checks}
\label{subsec:dataprepro}
\revise{We preprocessed raw EDA data with movisens DataAnalyzer~\cite{dataanalyzer}, which decomposes the signal into tonic SCL and phasic SCR components. It also reduces artifacts using a built-in low-pass filter ($<$10Hz), separates overlapping SCRs into distinct components~\cite{cowley2016short}, and uses sensor data such as temperature and activity to identify artifacts.}

\revise{The sensor exports aggregated per-second features, which we use directly without further downsampling. We use \textit{EdaScrAmplitudesMean} (\textmu S) as the phasic response magnitude $A(t)$ for all EDA-derived measures (Sec.~\ref{subsec:variables}), \textit{EdaScrRiseTimesMean} (s) for latency correction, and \textit{EdaSclMean} (\textmu S) as a tonic baseline for data-quality control. We treat SCL values below 0.3\textmu S as missing~\cite{stuldreher2025monitoring} and restrict computations to valid time windows. Absolute and relative timestamps are used to synchronize EDA and eye tracking and to compute viewing time.}

\subsection{Synchronization and Labeling using Eye-tracking Data}
\label{subsec:synchronization}
To compare story versions and arousal dynamics, we label each 1s EDA interval by story version, story order, and story piece using eye-tracking data to capture scroll-based transitions in the web-based slideshow.

\revise{First, we identify when a participant begins reading a story using the eye-tracking session video and map this relative time to the corresponding absolute timestamp in the eye-tracking data. We then manually align the eye-tracking and EDA recordings using absolute timestamps. Second, we extract scroll-transition timestamps from the screen recording. These define story-piece boundaries. Third, each EDA interval is assigned to the visible story piece. If participants navigate back and forth, story pieces can occur in multiple intervals, and all corresponding intervals are aggregated under the same story-piece label.}

\subsection{EDA-based Physiological Arousal Outcome Variables}
\label{subsec:variables}
We derive outcome variables from the SCR amplitude series for each participant and story view, reported in \textmu S unless noted otherwise. \revise{Because narrative responses may appear as sustained arousal or short, localized arousal peaks, we combine average, peak-sensitive, and segment-accumulation measures. This addresses the methodological gap that post-hoc self-reports and aggregate physiological measures may miss brief responses to specific visual-narrative moments.}

\noindent\textbf{Viewing time} represents the total duration that a participant viewed a story, derived from the eye-tracking-based segmentation (Sec.~\ref{subsec:synchronization}). \revise{We treat viewing time as a behavioral time-on-task measure, not as direct evidence of deeper emotional processing.}

\noindent\textbf{Average amplitude height (AvgAmpH)} per minute captures overall phasic intensity, normalized by viewing time: $AvgAmpH = \frac{\sum_{t \in T} A(t)}{\Delta(T)}$, where
A(t) is the SCR amplitude at second t, and $\Delta(T)$ is the valid duration in minutes.


\noindent\textbf{Maximum amplitude (MaxAmp)} captures the single highest SCR amplitude within the story interval:
$MaxAmp = \max_{t \in T} A(t)$. 

\noindent\textbf{Maximum sum of amplitudes (MaxSumAmp)} captures concentrated arousal within a story piece rather than a single spike. Let the story be segmented into K pieces with time intervals $T_1,...,T_K$. For each piece k, we compute: $SumAmp(k) = \sum_{t \in T_k} A(t)$.
We then take the maximum across pieces: $MaxSumAmp = \max_{k \in \{1,...,K\}} SumAmp(k)$.
\revise{This metric identifies the story piece with the strongest accumulated phasic activity. It is useful for slide-based stories but should be interpreted as arousal associated with a visible story piece, not as proof that this piece alone caused the response. Cumulative narrative context may also contribute.}

\subsection{Eye-tracking-based Peak Classification and Labeling}
\label{subsec:peak_classification}
To interpret which story elements are associated with strong physiological responses, we classify EDA peaks by the visual element viewed around the estimated eliciting moment. \revise{We include all detected SCR peaks for each participant and story view across the full story interval. Because SCR peaks lag behind eliciting stimuli, we estimate stimulus time by subtracting the per-second rise time and a typical SCR latency of 1.8s~\cite{edelberg1972electrodermal}:} $t_{stim}=t_{peak}-RiseTime(t_{peak})-1.8s$.
\revise{For each peak, we map the estimated stimulus time to the synchronized eye-tracking timeline, identify the nearest fixation sample, and label the fixation target as character illustration, data visualization, text, icon, or tooltip/quote area. Peaks without valid fixation data are labeled unassigned and excluded from element-frequency comparisons. This identifies elements associated with physiological arousal, but not whether the response was caused by the local image, the accumulated narrative arc, or both.}

\subsection{Questionnaire and Qualitative Feedback Analysis}
\revise{Quantitative questionnaire data included demographics (T1), emotion-selection after each story (T5, T7), user engagement after the first story (T5), and story preference after the second story (T7). Emotion-selection was repeated to maintain procedural symmetry and capture immediate affect before comparison. All questionnaires are provided in the supplemental material.}

\revise{The emotion-selection questionnaire used 20 labels to balance positive and negative affect with participant burden. Rather than capturing the full complexity of affective experience, it provided a structured self-report measure that could be compared across story pieces and related to EDA responses. We selected ten positive and ten negative emotions based on Cowen and Keltner's 27 emotions~\cite{cowen2017self}, excluding unlikely emotions and adding granularity for serious data stories. Participants selected one or more emotions for 14 representative story pieces across the five narrative phases. The main emotion was weighted as 2 and additional emotions as 1.}
User engagement was collected via 5-point Likert scales and summarized using mean scores for affective involvement, cognitive involvement, focused attention, usability, and novelty. This questionnaire design was adapted from Amini et al.~\cite{amini2018hooked} and O'Brien and Toms~\cite{obrien2009development}. \revise{Fig.~\ref{fig:results-user-engagement} reports the five category scores derived from the eight questionnaire items. Story preference was recorded only as a binary choice between the individual and general story version.}

\revise{Qualitative feedback came from open-ended questionnaire responses and optional informal comments during T8. The main study did not include recorded semi-structured interviews. We distinguish pilot think-alouds and interviews, which informed prototype refinement, from main-study free-text responses and T8 comments, which provided immediate and retrospective reflections. We analyzed the main study comments using exploratory, reflexive thematic analysis~\cite{braun2021thematic}. The first author iteratively open-coded written comments and notes, wrote analytic memos, and discussed emerging codes with the research team. Related codes were consolidated into four descriptive themes: empathy and personal connection; curiosity, surprise, and emotional reaction; learning and understanding; and cognitive load and information processing. Because this was lightweight feedback rather than dedicated interview data, we treat the themes as exploratory contextual evidence, not independent proof of the quantitative findings. The full thematic table is provided in the supplemental material.}

\subsection{EDA Data Statistical Analysis Workflow}
\revise{For story-level hypothesis testing, H1 and H2 use only the first story viewed to avoid order effects, while H3 uses both stories to compare the \revise{individual-first} and \revise{general-first} groups.} For each dependent variable and comparison, we assessed normality using the Shapiro--Wilk test, where $p<.05$ indicates deviation from normality. We used Welch's two-sample $t$-test for approximately normal variables because it is robust to unequal variances, and the Mann--Whitney $U$ test otherwise because it is robust to skew and outliers. Because H1--H3 are directional, we use one-tailed tests. For the three EDA-based arousal variables, we apply a Bonferroni correction~\cite{Haynes2013}, yielding $\alpha = .016$ ($.05/3$), and explicitly identify results that survive this threshold. All other analyses used $\alpha=.05$. \revise{For all hypothesis tests, we report $p$-values, effect sizes, and confidence intervals. Effect sizes quantify practical relevance: Cohen's $d$ for $t$-tests, Cliff's $\delta$ for Mann--Whitney $U$ tests, and risk ratio (RR) for binary peak-label comparisons. Confidence intervals are 98.4\% for Bonferroni-corrected EDA comparisons and 95\% otherwise.}


\section{Results}\label{sec:results} 
\revise{We report the results around RQ1--RQ2 and H1--H3 (Sec.~\ref{sec:experiment}). We first examine story-level physiological arousal and self-reported affect for the individual vs. general story (RQ1; H1--H2), then report eye-tracking-based peak attribution to contextualize arousal, and finally examine story-order effects (RQ2; H3) and self-reported user engagement.}

\subsection{Story-Level Physiological Arousal (RQ1/H1)}\label{sec:results-eda-story}

\revise{H1 posits greater physiological arousal for the individual-character story. We assess this using AvgAmpH, MaxAmp, and MaxSumAmp, with viewing time as an additional time-on-task measure.}

\begin{figure}[t]
  \centering

  \begin{subfigure}{\columnwidth}
    \centering
    \includegraphics[width=\columnwidth]{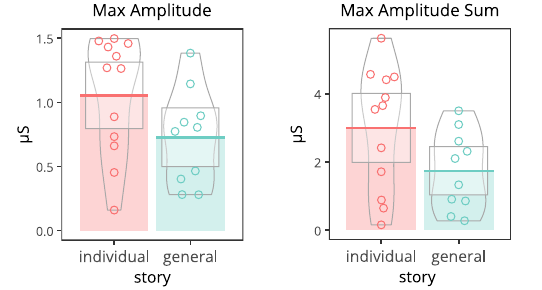}
        \vspace{-15px}
    \caption{}
     \label{fig:results-story-level-a}
  \end{subfigure}

  \vspace{-0.5em}

  \begin{subfigure}{\columnwidth}
    \centering
    \includegraphics[width=\columnwidth]{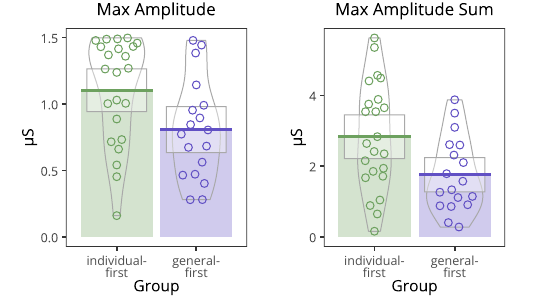}
    \vspace{-25px}
    \caption{}
     \label{fig:eda-groups}
  \end{subfigure}

  \caption{\revise{(a) Peak-based EDA measures by first viewed story version (individual vs. general). Each circle represents one participant. (b) Story-order effects comparing the individual-first and the general-first group.}}  
  \label{fig:results-story-level}
  \vspace{-0.8em}
\end{figure}

\revise{No significant differences were found for AvgAmpH. Peak-based measures were higher for the individual story (Fig.~\ref{fig:results-story-level-a}) but did not survive Bonferroni correction. MaxAmp showed a moderate effect size and a directional difference favoring the individual story (individual: \textit{M}=1.05, \textit{SD}=0.46; general: \textit{M}=0.73, \textit{SD}=0.37), but did not remain significant after correction (\textit{p}=.039 $>$ .016, \textit{d}=0.78, 98.4\% CI [-0.29, 1.85]). MaxSumAmp showed a similar pattern (individual: \textit{M}=3.00, \textit{SD}=1.79; general: \textit{M}=1.74, \textit{SD}=1.14; \textit{p}=.030 $>$ .016, \textit{d}=0.82, 98.4\% CI [-0.25, 1.90]). Viewing time was higher for the individual story (8:39 min vs. 7:21 min), but not significantly. We therefore interpret H1 as receiving only limited story-level EDA support.}

\subsection{Eye-Tracking Peak Attribution (RQ1 Complement)}
\revise{The peak analysis provides stronger element-level evidence. EDA peaks were significantly more associated with character illustrations in the individual story (\textit{M}=0.21, \textit{SD}=0.40) than with illustration elements in the general story (\textit{M}=0.10, \textit{SD}=0.29), \textit{p}=.001, RR=2.15, 95\% CI [1.34, 3.45]. Character illustrations accounted for 20.54\% of EDA peaks in the individual version and 9.54\% in the general version. This indicates an association between character-focused visual elements and physiological arousal, not a causal effect of the character element alone.}

\revise{Data visualizations were another most frequent peak source. Salient features in both stories included high-contrast dark-background scenes, staged iceberg reveals, and compositions guiding attention from characters or icons to symptom labels and prevalence curves. We call this compositional guidance within a story piece ``visual flow.'' Free-text and informal comments also highlighted character illustrations, while one participant preferred the general story's non-personalized style.}

\subsection{Self-Reported Emotional Profiles (RQ1/H2)}\label{sec:results-emotion-types}
\begin{figure}[t]
  \centering 
  \includegraphics[width=1.02\columnwidth, alt={Self-reporting on 20 emotion types for 14 story pieces.}]{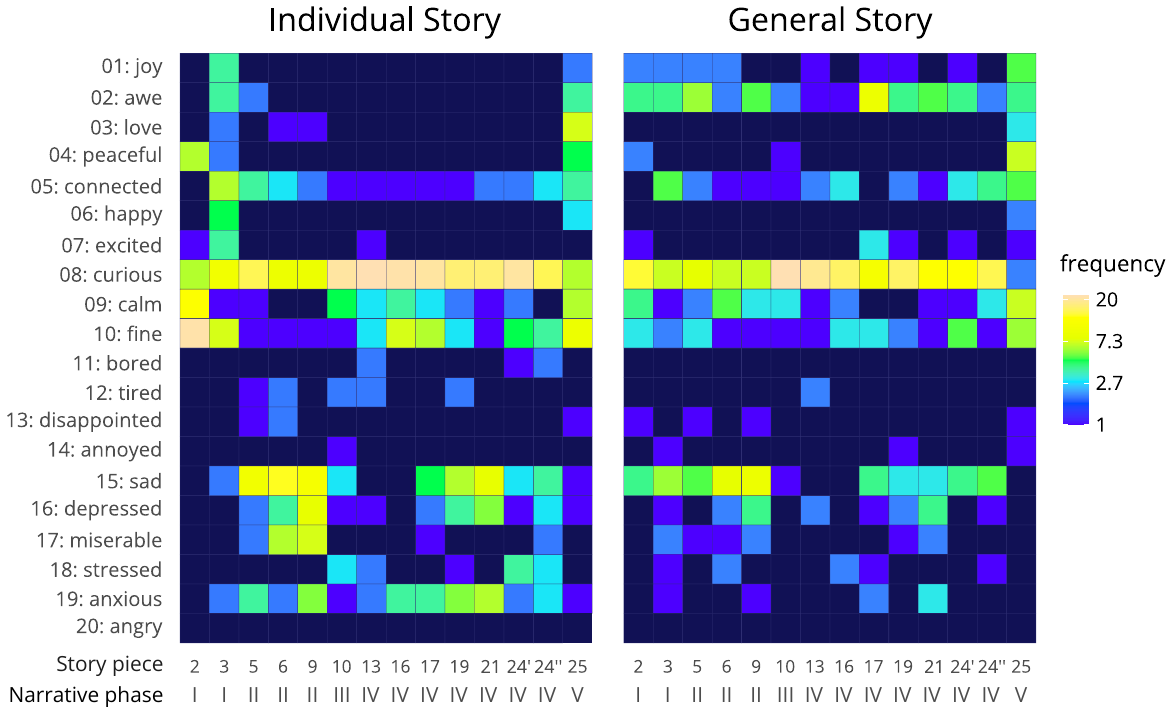}
  \caption{\revise{Distribution of 20 emotions across 14 selected story pieces (individual vs. general); color shows how many participants selected each emotion. X-axis labels denote narrative story-piece identifiers. Roman numerals indicate narrative phases: I Introduction, II Symptom Escalation, III Disease Explanation, IV Data-Driven Insight, and V Resolution. A detailed mapping of story-piece identifiers to narrative function and primary visual elements is provided in the supplemental material.}}
  \label{fig:emotions-matrix}
   \vspace{-0.5em}
\end{figure}

\revise{We report self-reported emotions for the first story viewed. Fig.~\ref{fig:emotions-matrix} shows 20 emotion labels across 14 rated story pieces. All except \textit{angry} were selected at least once. Since these pieces map to the shared narrative arc, the matrix complements the EDA peak analysis descriptively, without implying that phases alone caused physiological responses. Multiple emotions were reported more often for the individual story (64.5\%) than for the general story (49.0\%). Across both versions, \textit{curious} was the most frequent emotion and was higher for the individual story (\textit{M}=16.78, \textit{SD}=6.58) than for the general story (\textit{M}=12.14, \textit{SD}=5.46), \textit{p}=.029, \textit{d}=0.42, 95\% CI [-0.03, 0.73]. Because the confidence interval overlaps zero, we interpret this as a suggestive cognitive-affective response rather than strong emotion-specific evidence.}

\revise{The two stories also differed in positive emotion profiles. The general story produced significantly higher \textit{awestruck} responses (general: \textit{M}=3.79, \textit{SD}=2.15; individual: \textit{M}=0.71, \textit{SD}=1.49), \textit{p}=.00007, \textit{d}=0.80, 95\% CI [0.46, 0.94], and higher \textit{joy} responses (general: \textit{M}=1.21, \textit{SD}=1.37; individual: \textit{M}=0.43, \textit{SD}=1.16), \textit{p}=.011, \textit{d}=0.44, 95\% CI [0.03, 0.73]. Thus, the general story was not less affective, but seemed to support different emotional qualities, possibly related to scale, surprise, and collective relevance.}

H2 states that the individual story elicits stronger negative empathic emotions (\textit{sad}, \textit{depressed}, \textit{miserable}, \textit{anxious}) than the general story, assuming that character-driven narratives increase relatability and empathy.
We therefore examined these negative empathy-related emotions, which are linked to affective empathy in psychological research~\cite{malbois2023sympathy, goetz2010compassion, hatfield1993emotional}. \revise{They were significantly higher for the individual story (\textit{M}=3.04, \textit{SD}=3.45) than for the general story (\textit{M}=1.52, \textit{SD}=2.11), \textit{p}=.004, \textit{d}=0.27, 95\% CI [0.06, 0.46]. Thus, H2 is statistically supported, although the effect size is small.}

\revise{Comparing the heatmaps in Fig.~\ref{fig:emotions-matrix}, the individual story shows more variation across narrative phases, with positive emotions at the beginning and end, stronger negative empathy-related emotions around distressing pieces, and calmer middle responses. The general story shows a more uniform profile, except for the resolution and higher awe/joy responses. Thus, the framings seem to support different self-reported affective profiles rather than one story simply ``winning''. Together, these patterns suggest a possible ``Martini Glass'' structure for character-based medical stories, moving from a narrow individual anchor toward population-level evidence and multiple perspectives.}

\subsection{Story-Order Effects (RQ2/H3)}\label{sec:results-priming}
\revise{We evaluate H3 by analyzing effects of stimulus sequencing between both groups. We refer to these as story-order effects, because the design cannot fully disentangle emotional priming from repeated exposure, fatigue, novelty loss, interface learning, or cognitive effort.}
As shown in Fig.~\ref{fig:eda-groups}, \textit{MaxAmp} is significantly higher in the individual-first group (\textit{M} = 1.10, \textit{SD} = 0.39) than in the general-first group (\textit{M} = 0.81, \textit{SD} = 0.38), \textit{p} = .011, \textit{d} = 0.42, 98.4\,\% CI [0.05, 0.76], \revise{surviving correction.}
\textit{MaxSumAmp} showed the same pattern (individual-first: \textit{M}=2.83, \textit{SD}=1.51; general-first: \textit{M}=1.76, \textit{SD}=1.05), \textit{p}=.005, \textit{d}=0.83, 98.4\% CI [0.02, 1.60], also surviving correction with a moderate-to-large effect.
Viewing time was also higher in the individual-first group (\textit{M} = 453, \textit{SD} = 148) than in the general-first group (\textit{M} = 370, \textit{SD} = 157), \textit{p} = .045, \textit{d} = 0.56, 98.4\,\% CI [-0.22, 1.32], though this difference did not survive correction. The individual-first group spent 8:39 minutes on the first story and 6:22 minutes on the second; the general-first group spent 7:21 and 4:39 minutes, respectively.

\subsection{Self-Reported User Engagement}
\label{sec:results-engagement}

\begin{figure}[t]
  \centering 
  \includegraphics[width=\columnwidth, alt={Results on self-reported affective involvement, cognitive involvement, focused attention, usability, and novelty compared by story version.}]{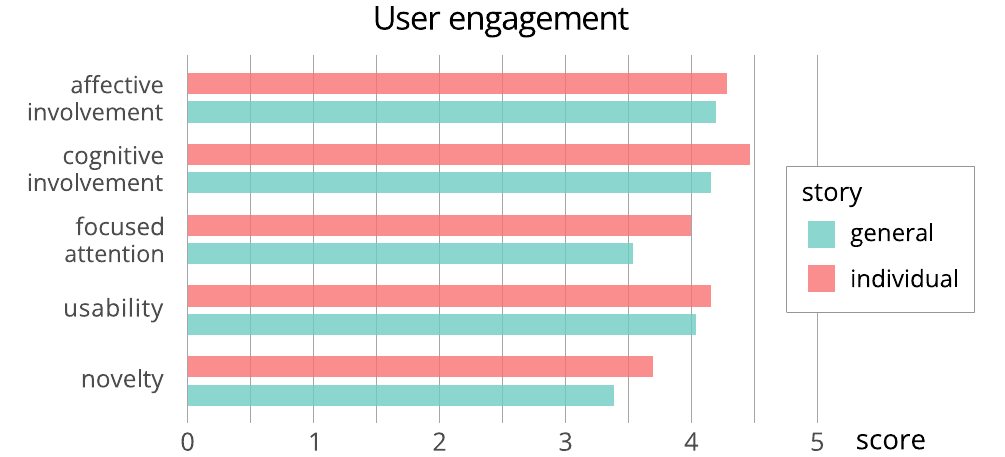}
 \caption{\revise{Self-reported user engagement by story version; eight questionnaire items aggregated into five categories: affective involvement, cognitive involvement, focused attention, usability, and novelty.}}
  \label{fig:results-user-engagement}
   \vspace{-0.5em}
\end{figure}

\revise{Finally, we report participants' self-reported user engagement after the first story view (see Fig.~\ref{fig:results-user-engagement}).} 
We found no significant difference in \textit{affective involvement}. Both versions elicited moderate connection to the affected women or the protagonist (Q1). 
Participants often expressed surprise at the disease's severity and the need for awareness (Q6--Q7).  \textit{Cognitive involvement} was similar for understanding (Q3), although the individual story was rated more insightful (Q5). \textit{Focused attention} was higher for the individual story (less mind-wandering, Q2). \textit{Usability} of interface elements was rated positively overall (Q8), with suggestions for clearer colors and interaction cues. \textit{Novelty} was high (Q4), despite some prior familiarity with the disease.

\section{Discussion}
\label{sec:disc-conc}

\revise{Our findings suggest that character-based and general population-level framings support different response profiles rather than a winner-loser pattern. We discuss these differences across physiological arousal, self-reported affective response, viewing time, and story-order effects.}

\subsection{What Do Characters Contribute? (RQ1)}
\revise{H1 received only limited story-level support. MaxAmp and MaxSumAmp showed moderate directional differences favoring the individual story, but neither survived Bonferroni correction, and both confidence intervals crossed zero; AvgAmpH showed no effect. We therefore conclude that the individual story did not produce stronger overall physiological arousal. Rather, it showed trend-level peak differences, while the strongest evidence comes from the element-level link between character illustrations and EDA peaks.}

\revise{The eye-tracking-based peak classification showed a robust association between character-focused elements and physiological arousal, suggesting that characters may act as local emotional anchors within a data story, especially at visually and narratively salient moments. However, EDA alone cannot explain why these moments were arousing. The self-reports contextualize this association: the individual story produced more negative empathy-related emotions and curiosity, suggesting that character-based framing may combine affective and cognitive-affective involvement.}

\revise{Viewing time should be interpreted cautiously. Participants spent more time with the individual story, but we did not ask whether this reflected emotional reflection. It may instead reflect character inspection, first-person narration, visual complexity, comprehension demands, curiosity, or cognitive effort. Since both stimuli were comparable but not perceptually equivalent, viewing time is best treated as a time-on-task indicator, not direct evidence of deeper emotional processing.}

\subsection{Emotional Affordances of Narrative Framing}
\revise{The general story's emotional profile is important: although the individual story elicited more negative empathy-related selections, the general story elicited significantly more awe and joy. This suggests that population-level framing can support valuable affective responses beyond personal empathy, possibly by emphasizing the scale, prevalence, and collective relevance of HG through the iceberg metaphor, population-style icons, and quotes from multiple women.}

\revise{This distinction matters for public health communication. Individual narratives may foster empathy, personal resonance, and concern for a patient's lived experience. Population-based narratives may better communicate scope, social relevance, or collective responsibility. Our results therefore suggest that character-based and general framings may be complementary rather than hierarchical.}

\subsection{Curiosity, Learning, and Multimodal Discrepancies}
\revise{Curiosity was the most frequent self-reported emotion across both stories, complicating a purely emotional interpretation of EDA. It may reflect novelty, interest in the visual explanation, surprise about disease severity, cognitive effort, or learning-oriented engagement with the data. Because we did not measure prior knowledge or learning outcomes, we cannot determine whether curiosity corresponded to actual learning. Future studies should distinguish novelty-driven from learning-oriented curiosity using knowledge gain, perceived insight, or delayed recall.}

\revise{Our findings show why physiological and self-report measures should be interpreted together. EDA provided time-resolved arousal indicators, while self-reports supplied emotion labels and valence, and the two did not always align. This is expected: EDA can capture arousal associated with emotional response, curiosity, cognitive effort, attention, or surprise. We therefore interpret emotional engagement only where arousal, gaze behavior, and self-reported emotion converge.}

\subsection{Methodological Implications for Peak-based Analysis}
\revise{Methodologically, the contrast between average and peak-sensitive EDA measures was informative. Average arousal did not distinguish story versions, while peak-sensitive measures showed directional differences and supported element-level interpretation. Thus, fine-grained, peak measures can reveal localized arousal patterns that aggregate measures may obscure, though this is a methodological insight rather than conclusive evidence that narrative structure caused the peaks.}

\revise{Stories build meaning over time, so peaks during a character illustration, iceberg metaphor, or data visualization may reflect the immediate visual stimulus, the cumulative narrative context established by preceding slides, or both. Because emotion labels were collected only for selected story pieces, we interpret peak-based results as arousal associated with specific visual-narrative moments, not proof that isolated images or elements independently caused the response.
}

\subsection{Preliminary Story-Order Effects (RQ2)}
\label{subsec:disc-priming}
\revise{The H3 results indicate a story-order effect: the individual-first group showed higher peak-based arousal than the general-first group. Character-based emotional priming is plausible, but the design cannot isolate it from fatigue, reduced novelty, repeated HG exposure, learning interaction, or cognitive effort from the longer individual story. We therefore describe this as preliminary evidence of an order-dependent response, not definitive evidence of personalization-induced priming.}
\revise{To our knowledge, such order-dependent arousal has not been reported in prior narrative medical visualization studies. While intriguing, broader conclusions require replication with larger samples, topics, and stimulus pairs. Future studies could add pre-story baselines, fatigue or cognitive-load ratings, rest or filler tasks, and between-subject designs.}

\subsection{Design Implications and the Martini Glass Hypothesis}
\revise{The findings suggest preliminary design implications. Designers may introduce an individual character early to establish relevance or empathy, without making the character dominate the whole story. Character-focused scenes may work best at emotional moments, while explanatory visualizations can carry the broader data argument. Explicit transitions from individual experience to population-level relevance can help preserve emotional anchoring when shifting to aggregate data.}

\revise{A worked HG example would start with Freja’s experience, use her symptoms to motivate the data, transition to the iceberg visualization to reveal hidden symptoms, and then broaden through prevalence curves and quotes from multiple women. Other public-health stories could follow a similar arc: introduce one patient, show lived consequences, transition to prevalence statistics, present broader population trends, and close with societal implications and action.}
\revise{We therefore frame this ``Martini Glass'' structure as a preliminary design hypothesis, not a validated guideline. We hypothesize that a story may benefit from opening with a narrow individual anchor and gradually broadening toward population-level evidence and multiple perspectives. This is informed by the story-order analysis, character-related EDA peaks, empathy-related self-reports, and the general story's awe/joy profile, but requires direct testing against alternative structures.}

\subsection{Limitations} 
We outline the study's main limitations and discuss factors that may affect the interpretation and generalizability of the results.

\noindent\textbf{Small sample size and statistical power:} \revise{The small sample (N=26) and between-subjects order comparison limited statistical power. For H1, peak-based EDA effects were moderate but did not survive correction, with confidence intervals that crossed zero. So claims about character-based emotional engagement rely on convergent evidence from eye-tracking peak attribution, self-reported emotions, and descriptive trends rather than story-level EDA significance alone.}

\noindent\textbf{Causal interpretation:} \revise{Although topic, sequence, visualizations, and interaction structure were closely matched, the stories differed in visual and rhetorical presentation. 
Arousal or viewing-time differences, therefore, cannot be attributed to character presence alone with certainty.}

\noindent\textbf{Story-order confounds:} \revise{The sequential design counterbalanced order but cannot separate personalization from carryover, fatigue, reduced novelty, learning, or cognitive effort. H3 should therefore be read as a preliminary order effect, not definitive evidence of emotional priming.}

\noindent\textbf{Experimental setup:} Physiological measurements require participant contact, making them costlier and slower than surveys and contributing to our small sample. \revise{EDA captures time-resolved arousal, but it is not more truthful than self-report and needs contextual interpretation.}

\noindent\textbf{Generalizability:} \revise{Most participants held a university degree, which may limit generalizability to broader publics. However, we did not assess visualization, statistical, or health literacy, and therefore do not infer data-interpretation ability from background. Emotional responses may also depend on prior health and pregnancy experience, HG familiarity, caregiving roles, and broader biopsychosocial context. The excluded former HG patient illustrates this: her unusually strong EDA responses precluded group comparisons, but may indicate that lived experience shapes physiological responses to medical narratives.}

\noindent\textbf{Bias of topic and design:} The specific topic and story design also limit generalization. As Mittenentzwei et al.~\cite{mittenentzwei2023disease} proposed, testing the same hypotheses across disease topics would improve generalizability, but reduce personalization through character experience. Visual choices such as color palette or illustration style may also affect perception. \revise{Thus, our design implications are preliminary and may not generalize across audiences with different health histories, cultural contexts, or relationships to pregnancy-related illness.}

\noindent\textbf{Synchronization and gaze interpretation:} Eye-tracking and EDA were aligned manually using story-start events and absolute timestamps. Although both systems provide high-resolution timestamps, no drift correction was applied, so minor timing errors may affect peak-to-event assignments. \revise{Gaze is also not a direct proxy for affective involvement: participants may look away from distressing or personally relevant content, so avoidance can signal emotional response rather than disengagement. Our peak attribution may therefore miss moments when participants looked away from the screen or a relevant element. We also did not fully analyze gaze, such as dwell time, fixation counts, or transitions, which future work could use to study visual processing.}

\noindent\textbf{Measurements:} EDA measures include uncertainty: automated SCR detection sensitivity has an accuracy of 92\%, and the EdaMove 4 sensor slightly underestimates large peaks ($>$2\textmu S) by 5--10\% due to 14-bit quantization and 32Hz sampling, although our analysis software attempts correction. The delay between a stimulus and the physical SCR varies among individuals and has been estimated from prior work~\cite{edelberg1972electrodermal}. Furthermore, respiration can induce EDA peaks, but was not recorded. Instead, we focused on maximum EDA amplitudes to reduce this issue. \revise{ Finally, EDA may reflect cognitive involvement, curiosity, attention, or learning, and cannot distinguish between positive and negative valence, necessitating other data streams for interpretation.}

\noindent\textbf{Emotion and self-report constraints:} 
Social desirability may affect emotion selection, as participants may avoid reporting socially undesirable reactions. \revise{The set of 20 emotion labels reduced burden and enabled comparison, but simplified affective experience and may miss mixed or nuanced emotions. We therefore interpret the emotion results as indicators of affective tendencies, not a comprehensive account of emotional experience. Qualitative feedback was also lightweight rather than based on semi-structured interviews. Although we added an exploratory reflexive thematic analysis~\cite{braun2021thematic}, the themes provide contextual interpretations of open-ended feedback, not a full explanation of the EDA and eye-tracking patterns. Future work should use richer qualitative protocols to examine why story elements elicit awe, joy, curiosity, empathy, or fatigue, and could add physiological measures, such as electroencephalography or functional magnetic resonance imaging, to deepen insight into emotional and cognitive processes~\cite{samur2024getting}.}

\section{Conclusion \& Future Work}\label{sec:future}
\revise{This work offers empirical and methodological insights into narrative medical visualization by combining physiological arousal, eye-tracking peak attribution, self-reported emotions, and lightweight qualitative feedback. Rather than showing a uniform benefit of character-driven stories, our results reveal differentiated profiles. The individual story showed limited story-level EDA support, but character illustrations were strongly linked to EDA peaks and elicited more negative empathy-related emotions. The general story, in contrast, elicited more awe and joy, suggesting value for population-level public health communication.
Methodologically, combining average and peak-sensitive EDA measures with eye tracking revealed localized arousal patterns that averages could obscure, while self-reports clarified their affective quality.}

\revise{Future work should replicate the study with larger, more diverse samples, additional medical topics, and multiple stimulus pairs. It should also measure visualization and health literacy, fatigue, cognitive load, prior topic familiarity, and learning outcomes, and include larger subgroups with lived experience, caregivers, clinicians, and public audiences. 
Finally, the proposed Martini Glass structure should be tested by comparing character-first, population-first, and hybrid structures.}

\section*{Supplemental Materials}\label{sec:supps}
\label{sec:supplemental_materials}

The supplementary material is available via PCS and OSF at \url{https://doi.org/10.17605/OSF.IO/NXBY3}, released under a CC BY 4.0 license. It includes the appendix, full PDF versions of both stories, a mapping of story pieces to narrative phases and visual elements, all questionnaires and corresponding results, including the full coding table and participant responses, recorded and cleaned EDA data, a link to Python scripts for data analysis, and GitHub links to the interactive story implementations. Due to the file size limit in PCS, the questionnaire files uploaded there contain reduced-resolution images. High-resolution versions of the questionnaires, including the “Emotion Selection” questionnaire, are available in the OSF repository.

\acknowledgments{%
MoBa: We thank the Norwegian Institute of Public Health (NIPH) for gathering and maintaining the MoBa data set. All analyses in the MoBa cohort were performed using digital labs in HUNT Cloud at the Norwegian University of Science and Technology, Trondheim, Norway. 
Marc Vaudel was supported by the Research Council of Norway (\#301178), the European Research Council (\#101171420), and the University of Bergen.  
}

\bibliographystyle{abbrv-doi-hyperref-narrow}

\bibliography{template}

\clearpage
\newpage

\appendix 

\begin{figure*}[h!]
  \centering
  \begin{subfigure}{0.49\textwidth}
    \centering
    \includegraphics[width=\linewidth]{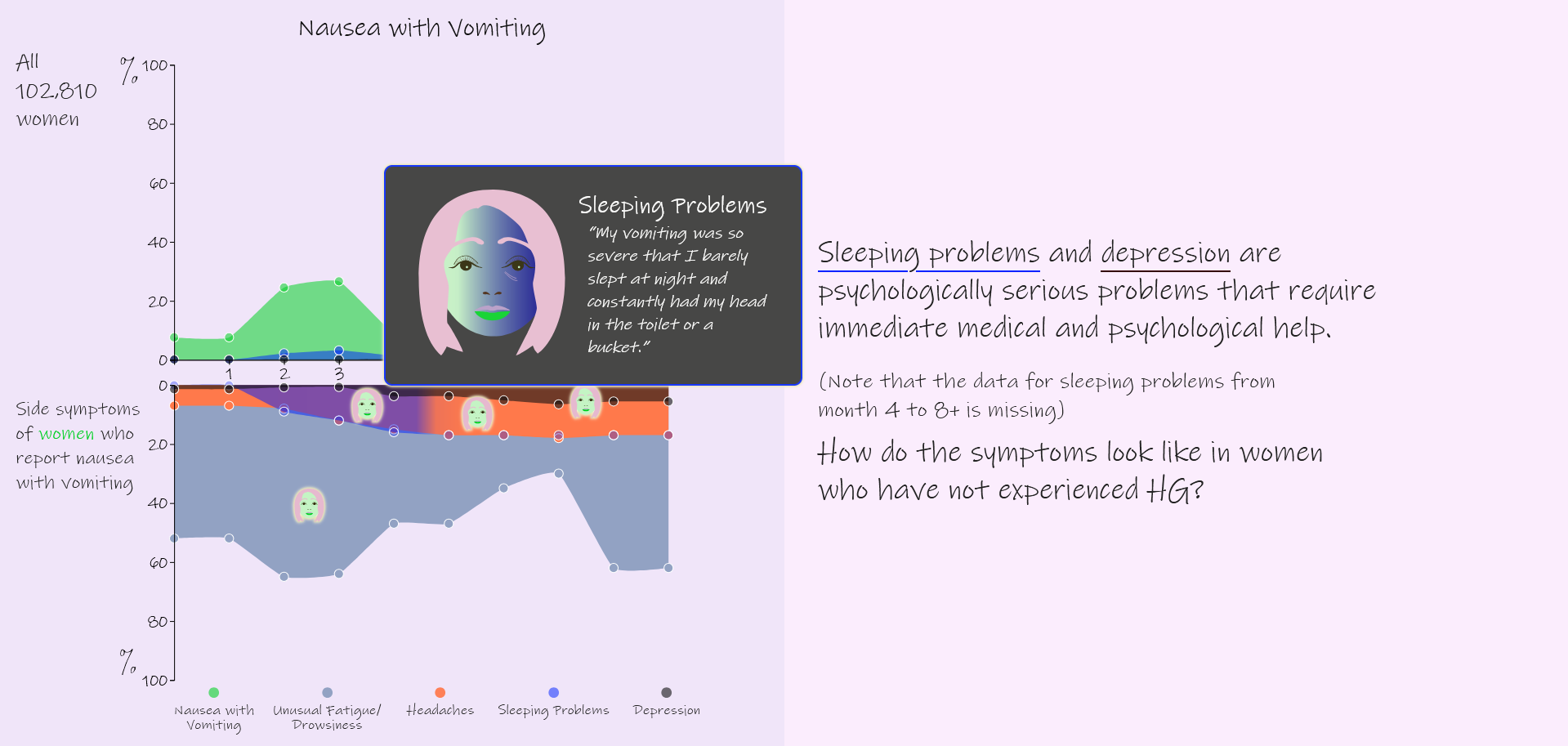}
    \caption{}
    \label{fig:ind-a}
  \end{subfigure}
  \hfill
  \begin{subfigure}{0.49\textwidth}
    \centering
    \includegraphics[width=\linewidth]{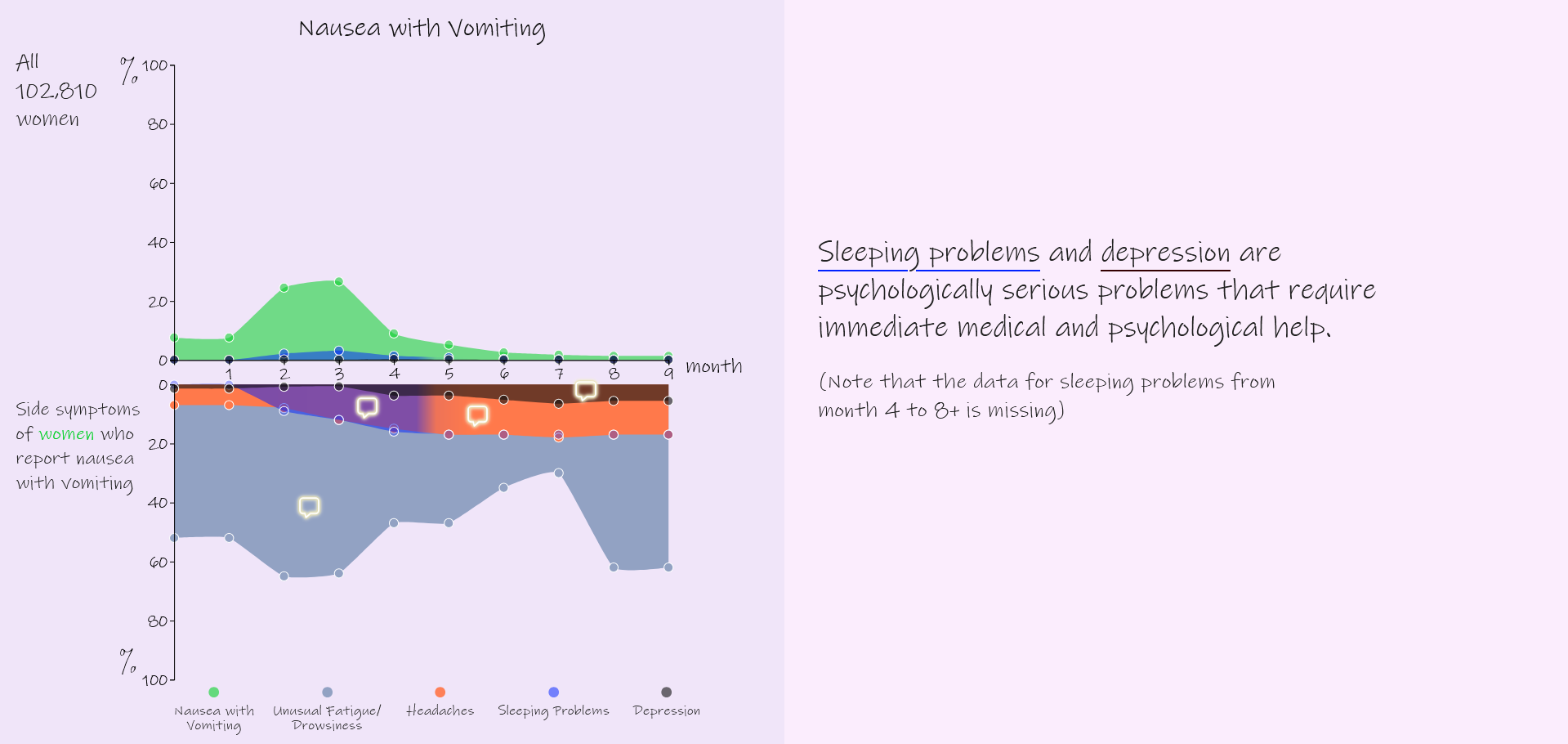}
    \caption{}
    \label{fig:ind-b}
  \end{subfigure}
  \caption{Exemplary story piece from the \emph{Data-Driven Insights} phase showing the same MoBa-based symptom visualization in the individual and general story versions. (a) The individual story includes character-linked cues and a character-associated tooltip. (b) The general story uses the same chart structure and data but replaces the character framing with icon-style markers and non-personalized contextual text.}
  \label{fig:results-story-level_1}
\end{figure*}

\begin{figure*}[ht!]
  \centering
  \begin{subfigure}{0.49\textwidth}
    \centering
    \includegraphics[width=\linewidth]{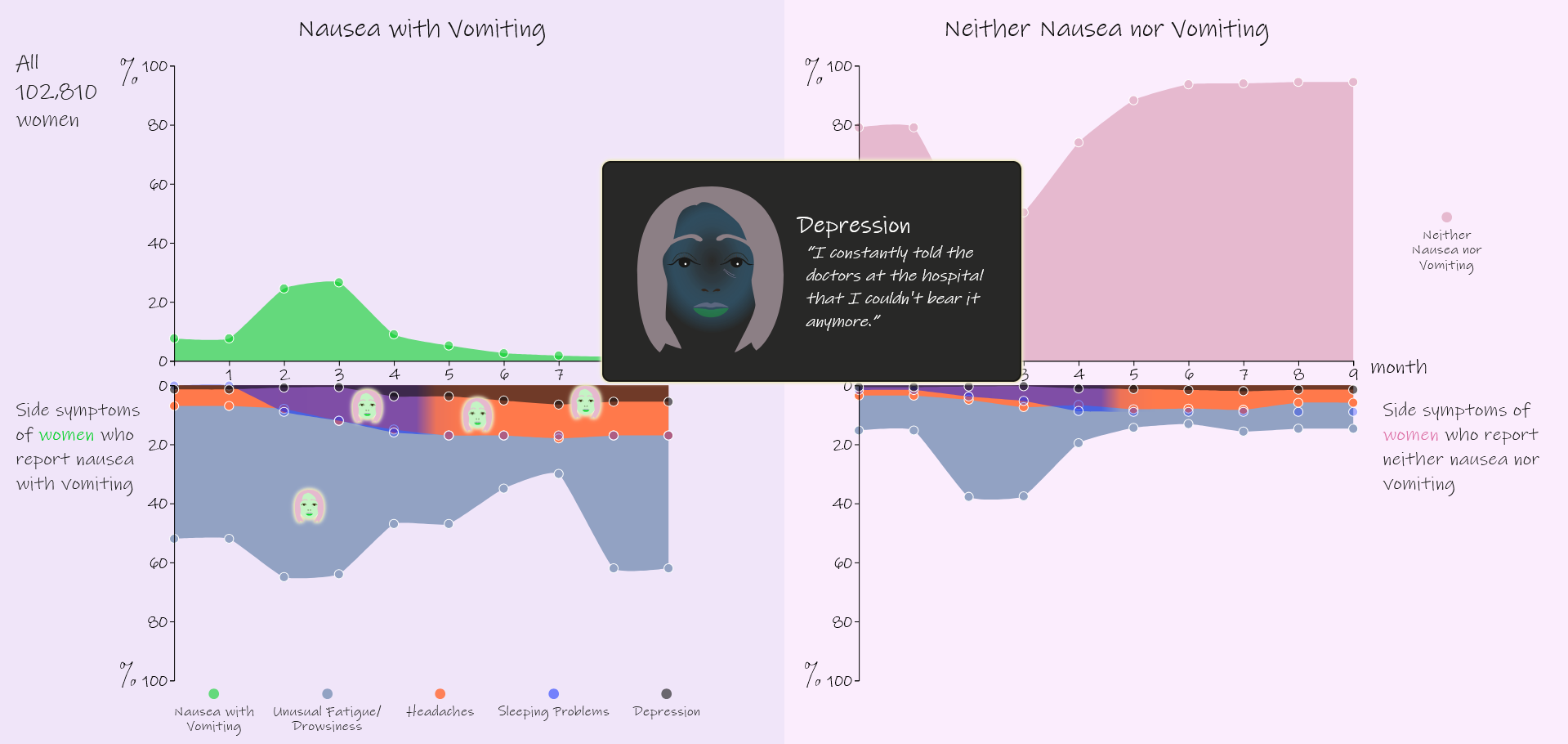}
    \caption{}
    \label{fig:gen-a}
  \end{subfigure}
  \hfill
  \begin{subfigure}{0.49\textwidth}
    \centering
    \includegraphics[width=\linewidth]{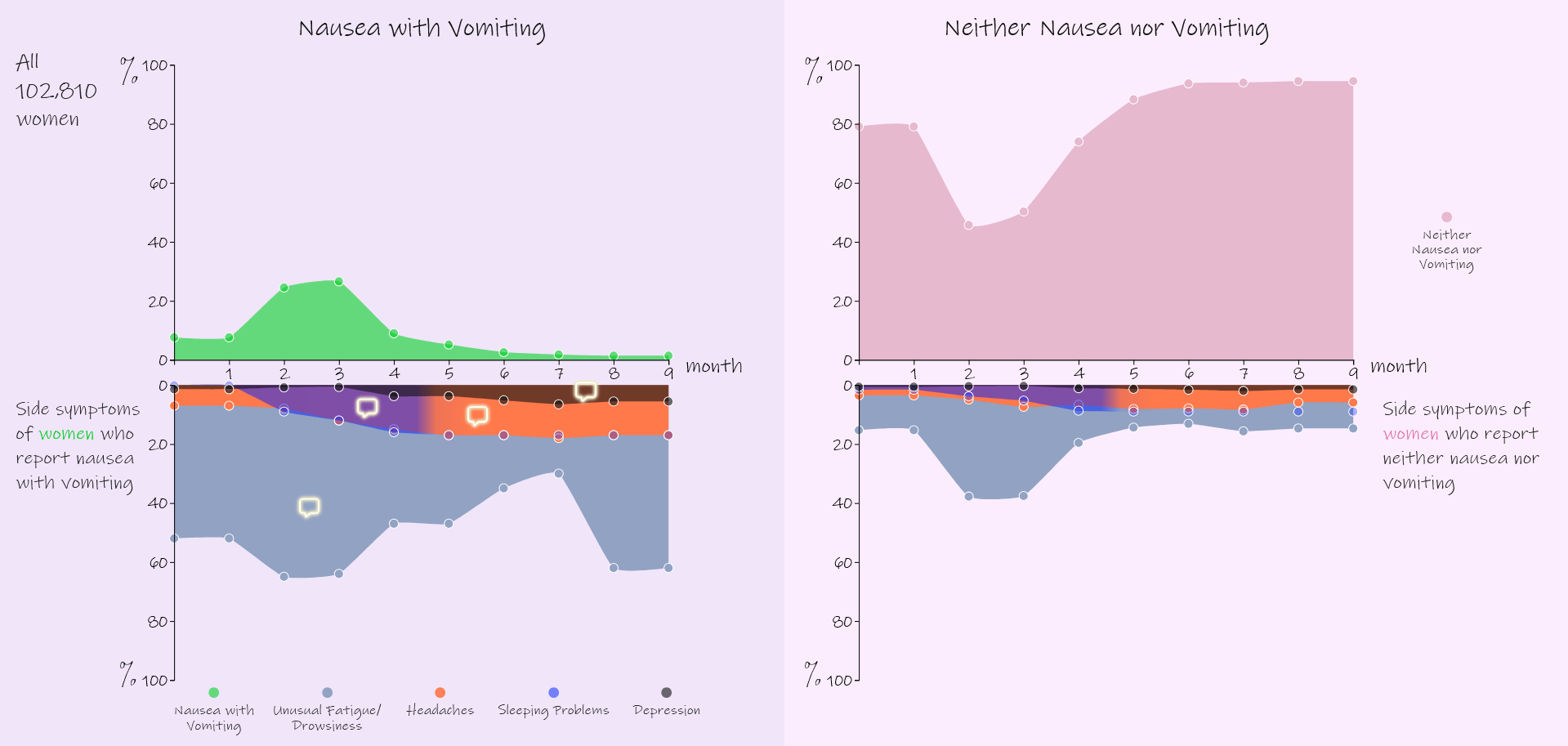}
    \caption{}
    \label{fig:gen-b}
  \end{subfigure}
  \caption{Exemplary comparative story piece from the \emph{Data-Driven Insights} phase. Both versions present the same MoBa-based comparison between women with nausea and vomiting and those with neither. (a) The individual story uses character-linked markers and a character-associated tooltip. (b) The general story preserves the same chart structure and data but uses icon-style markers and non-personalized contextual text.}
  \label{fig:results-story-level_2}
\end{figure*}

\section{MoBa Data}
\label{app:Moba}
\subsection{MoBa Data Availability}
Data from the Norwegian Mother, Father and Child Cohort Study used in this study are managed by the Norwegian Institute of Public Health and can be made available to researchers, provided approval from the Regional Committees for Medical and Health Research Ethics (REC), compliance with the EU General Data Protection Regulation (GDPR), and approval from the data owners. The consent given by the participants does not allow for the storage of data on an individual level in repositories or journals. Researchers who want access to data sets for replication should apply through helsedata.no. Access to data sets requires approval from the Regional Committee for Medical and Health Research Ethics in Norway and an agreement with MoBa.

\subsection{Ethics Declarations}
MoBa: Informed consent was obtained from all study participants. The administrative board of the Norwegian Mother, Father and Child Cohort Study, led by the Norwegian Institute of Public Health, approved the study protocol. The establishment of MoBa and initial data collection was based on a license from the Norwegian Data Protection Agency and approval from the Regional Committee for Medical Research Ethics. The MoBa cohort is currently regulated by the Norwegian Health Registry Act. The study was approved by the Regional Committee for Medical Research Ethics (744956). No specific ethical approval was required for the character-based visualization study according to institutional guidelines. Our study adhered to the Declaration of Helsinki. Written informed consent was obtained from all participants, and participant privacy was preserved by anonymizing the study data.

\subsection{MoBa Variables and Preparation}
Maternal symptom prevalence stratified by week of pregnancy was obtained from answers to questionnaires 1 and 3 returned by mothers in MoBa. Symptom prevalence was stratified by NVP status according to the answer to questionnaires: (No NVP) the mother did not report nausea or vomiting at any time point; (NVP) the mother reported nausea or vomiting at a given time point; (HG) the mother reported hospitalization during pregnancy due to prolonged nausea or vomiting.

\section{Exemplary Story Screenshots}
\label{app:story-screenshots}

Larger exemplary screenshots of selected story pieces are provided in Figure~\ref{fig:results-story-level_1} and Figure~\ref{fig:results-story-level_2} to illustrate the visual and narrative differences between the individual and general story versions.

\section{Additional Data Analysis Details}
\label{app:data-analysis-details}

This appendix complements Sec.~\ref{sec:data-analysis} of the main manuscript with additional implementation details from the data collection and analysis workflow. The main manuscript describes the overall workflow and defines the EDA-derived outcome variables; here, we provide further information on exported EDA features, synchronization, story-piece labeling, peak-to-gaze mapping, and the emotion-selection questionnaire.

\subsection{Additional EDA Export and Validity Details}
\label{app:eda-export-quality}

This appendix adds implementation details to the preprocessing summary in Sec.~\ref{subsec:dataprepro}. We distinguish between \emph{exported per-second features}, i.e., DataAnalyzer summaries for each 1\,s window, and the derived outcome variables used for hypothesis testing in Sec.~\ref{subsec:variables}. The amplitude-related export underlies the physiological arousal metrics, while rise-time and SCL features support latency correction and data-quality control.

The sensor provides 14-bit EDA resolution, and all analyses were conducted at 1\,s resolution, corresponding to one row per second. We used the exported features directly without downsampling: \textit{EdaScrAmplitudesMean} (\textmu S) serves as $A(t)$ for the EDA-derived measures (Sec.~\ref{subsec:variables}), \textit{EdaScrRiseTimesMean} (s) supports latency correction, and \textit{EdaSclMean} (\textmu S) provides the tonic baseline for validity checks. Following prior work~\cite{stuldreher2025monitoring}, computations were restricted to valid time windows with $\mathrm{SCL} \geq 0.3\,$\textmu $S$.

\subsection{Additional Synchronization and Labeling Details}
\label{app:sync-labeling}

This appendix provides implementation-level details for the synchronization and labeling procedure summarized in Sec.~\ref{subsec:synchronization}.

\noindent\textbf{Time bases and alignment.}
The eye-tracking session video uses a relative timeline starting at 00{:}00{:}00, whereas the eye-tracking export provides absolute timestamps with microsecond precision. For each participant, we identified the story-start moment in the video based on the visible story content and mapped this relative time to the corresponding absolute timestamp in the eye-tracking export. We then aligned the EDA and eye-tracking recordings using the first absolute timestamp available in both recordings. This established a synchronized time base for slicing the EDA signal. Because both streams were timestamped during acquisition, no additional drift correction was applied.

\noindent\textbf{Scroll-transition representation.}
Scroll transitions in the screen recording were converted to absolute timestamps using the synchronized time base. We represent the resulting sequence of story-piece boundaries as: $B = \{(b_i, p_i)\}_{i=1}^{m}$,
where $b_i$ denotes the transition time, $p_i$ the story piece shown after the transition, and $m$ the total number of recorded scroll transitions. Back-and-forth navigation was retained rather than simplified, so the same story piece could occur in multiple disjoint intervals.

\noindent\textbf{Implementation of interval labeling.}
We implemented the labeling procedure in Python using pandas and numpy. For each participant, the 1\,Hz EDA time series was segmented into story-level and story-piece-level intervals based on synchronized absolute timestamps. Each 1\,s EDA interval was assigned to the currently visible story piece by propagating the most recent scroll-transition label.

Each labeled interval contains the fields \{participant\_id, group, story\_version, story\_view, story\_order, story\_piece\_id\}. If a story piece was revisited, all corresponding intervals were assigned the same story\_piece\_id. Thus, piece-level computations aggregate across all visits to that story piece, including non-contiguous intervals.

\subsection{Additional Notes on EDA-Derived Variables}
\label{app:eda-variable-notes}

The main manuscript defines the EDA-derived variables AvgAmpH, MaxAmp, and MaxSumAmp (Sec.~\ref{subsec:variables}). Here, we provide the rationale for using these complementary variables.

Narrative responses can appear as sustained involvement, prolonged exposure, or punctuated peaks of arousal. We therefore combine duration-normalized, peak-sensitive, and segment-accumulation measures~\cite{boucsein2012electrodermal, van2012emotional, benedek2010, saket2016}. Prior work shows that mean and maximum EDA features can capture stimulus-induced emotional changes~\cite{jukiewicz2025analysis}. We use mean-based measures to represent overall phasic arousal, maximum-based measures to capture the strongest momentary response, and segment-accumulation measures to identify story pieces with repeated or sustained phasic activity.

Viewing time is the total duration, in seconds, that a participant viewed a story, derived from the eye-tracking-based segmentation described in App.~\ref{app:sync-labeling}. In visualization research, time-on-task is often used as an indicator of behavioral engagement~\cite {saket2016}. In this study, we interpret it more conservatively as behavioral exposure to the story.

For MaxSumAmp, the accumulated amplitude within a story piece approximates integrated phasic activity during that interval and follows prior work recommending time integration of EDA to capture sympathetic activation, particularly when responses overlap~\cite{benedek2010}. The measure is well-suited for slide-based stories, where arousal may cluster around specific segments such as character-focused scenes, staged data revelations, or metaphorical visuals.

\subsection{Additional Peak Classification Details}
\label{app:peak-classification}

This appendix adds implementation details to the peak-classification procedure summarized in Sec.~\ref{subsec:peak_classification}. The main manuscript describes the latency correction, fixation-target categories, and interpretive limitations.

\noindent\textbf{Peak selection.}
Candidate peaks were derived from the SCR amplitude time series. We included all detected SCR peaks for each participant and story view across the complete story interval, rather than selecting only the largest peaks. This preserves the full range of phasic responses and avoids bias toward unusually salient events. SCR peaks are interpreted as phasic sympathetic activation~\cite{boucsein2012electrodermal}.

\noindent\textbf{Operationalizing latency correction.}
For each peak, the per-second mean rise time was taken from \textit{EdaScrRiseTimesMean} at the peak time point. This value was combined with the fixed 1.8\,s latency term described in the main manuscript to estimate the likely stimulus time. This estimated stimulus time was then mapped to the synchronized eye-tracking timeline.

\noindent\textbf{Mapping peaks to gaze.}
For each estimated stimulus time, we selected the nearest available fixation sample in time and used its fixation coordinates $(x,y)$ to determine the viewed story element. Labels were assigned based on the fixation target visible in the synchronized screen recording. If no valid fixation was available, for example, due to missing gaze data, the peak was labeled as unassigned and excluded from element-frequency comparisons.

\noindent\textbf{Peak-level records.}
The resulting peak-level records contain the following fields:
\{participant\_id, story\_version, story\_piece\_id, peak\_rank, peak\_amplitude, story\_element\_label, story\_element\_category\}.
These records form the basis for the element-frequency comparisons reported in Sec.~\ref{sec:results} of the main manuscript.

\subsection{Details of the Emotion-selection Questionnaire}
\label{app:questionnaire-handling}

In the emotion-selection questionnaires, we provided 20 emotion labels supported by a graphical representation. The emotion set was based on Cowen and Keltner's list of 27 emotions~\cite{cowen2017self}. We excluded emotions that we did not expect to occur in this context, such as \textit{romance} or \textit{horror}, and added more granularity for categories expected to be relevant in serious data stories. For example, sadness was represented through \textit{sad}, \textit{depressed}, and \textit{miserable}.

To ensure balance, we selected ten positive and ten negative emotions covering a range of arousal levels from low to high. Following Ekman~\cite{ekman1999basic}, positive emotions were ordered by decreasing positivity and intensity: \textit{joy}, \textit{awe}, \textit{love}, \textit{peaceful}, \textit{connected}, \textit{happy}, \textit{excited}, \textit{curious}, \textit{calm}, and \textit{fine}. Negative emotions were ordered by increasing negativity and intensity: \textit{bored}, \textit{tired}, \textit{disappointed}, \textit{annoyed}, \textit{sad}, \textit{depressed}, \textit{miserable}, \textit{anxious}, \textit{stressed}, and \textit{anger}.

\begin{figure}[ht]
  \centering

  \begin{subfigure}{\columnwidth}
    \centering
    \includegraphics[width=\columnwidth]{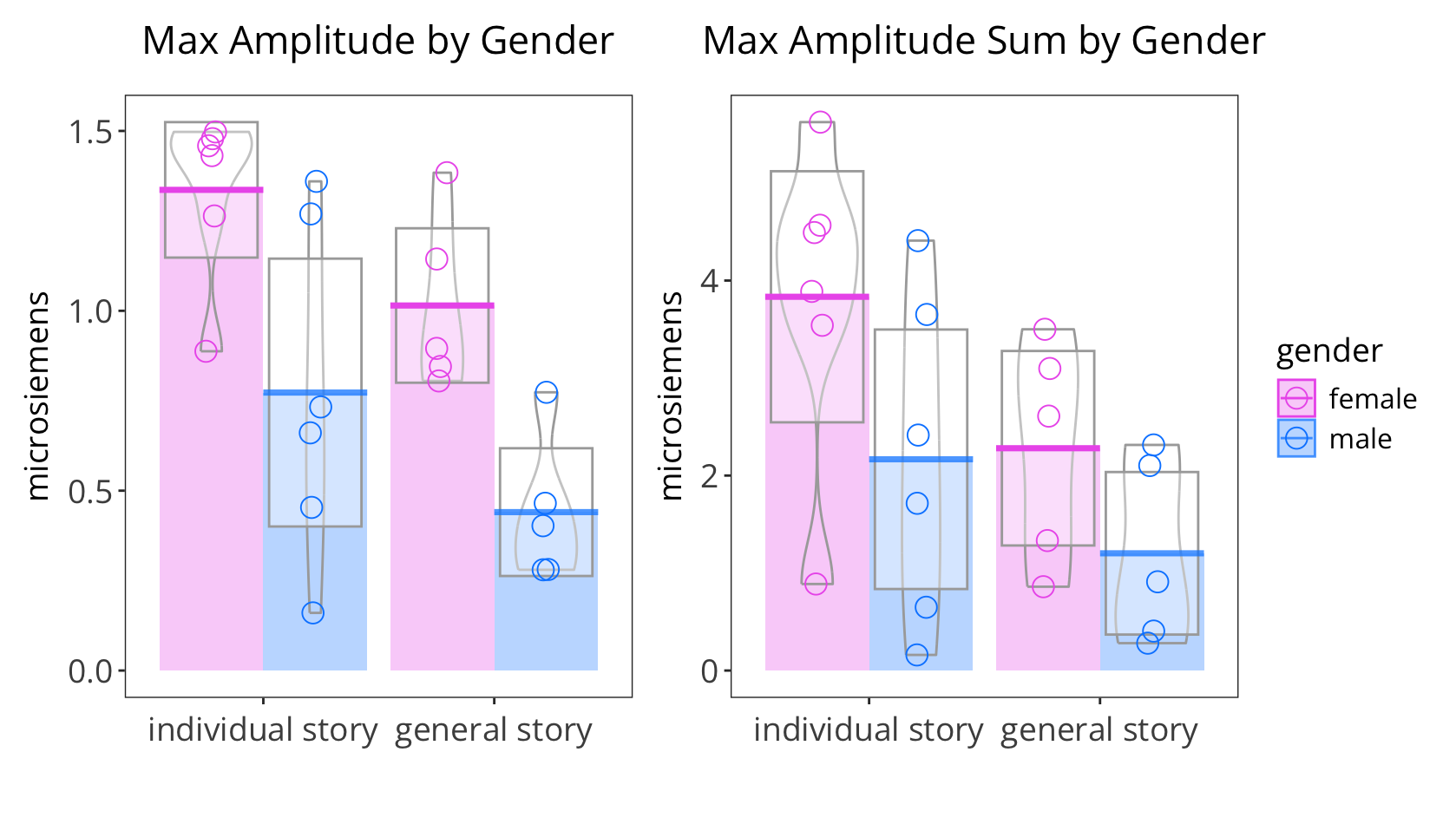}
        \vspace{-25px}
    \caption{}
     \label{fig:veri-a}
  \end{subfigure}

  \vspace{-0.5em}

  \begin{subfigure}{\columnwidth}
    \centering
    \includegraphics[width=\columnwidth]{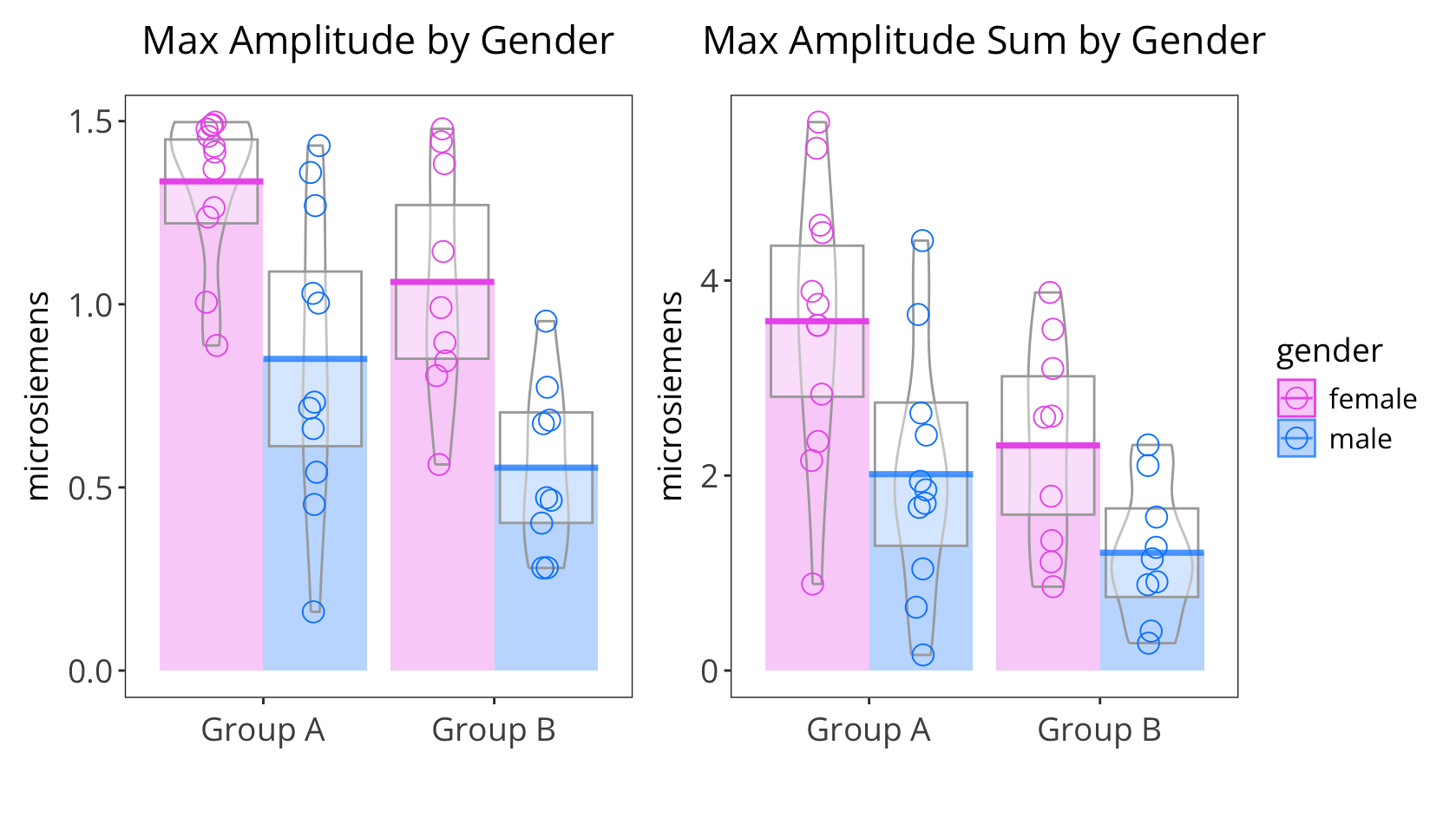}
    \vspace{-25px}
    \caption{}
     \label{fig:veri-b}
  \end{subfigure}

  \caption{Large differences in EDA measures between genders, but main effects for story version and among participant groups were maintained.}
  \label{fig:veri}
\end{figure}

\begin{figure}[ht]
    \centering
    \includegraphics[width=\columnwidth]{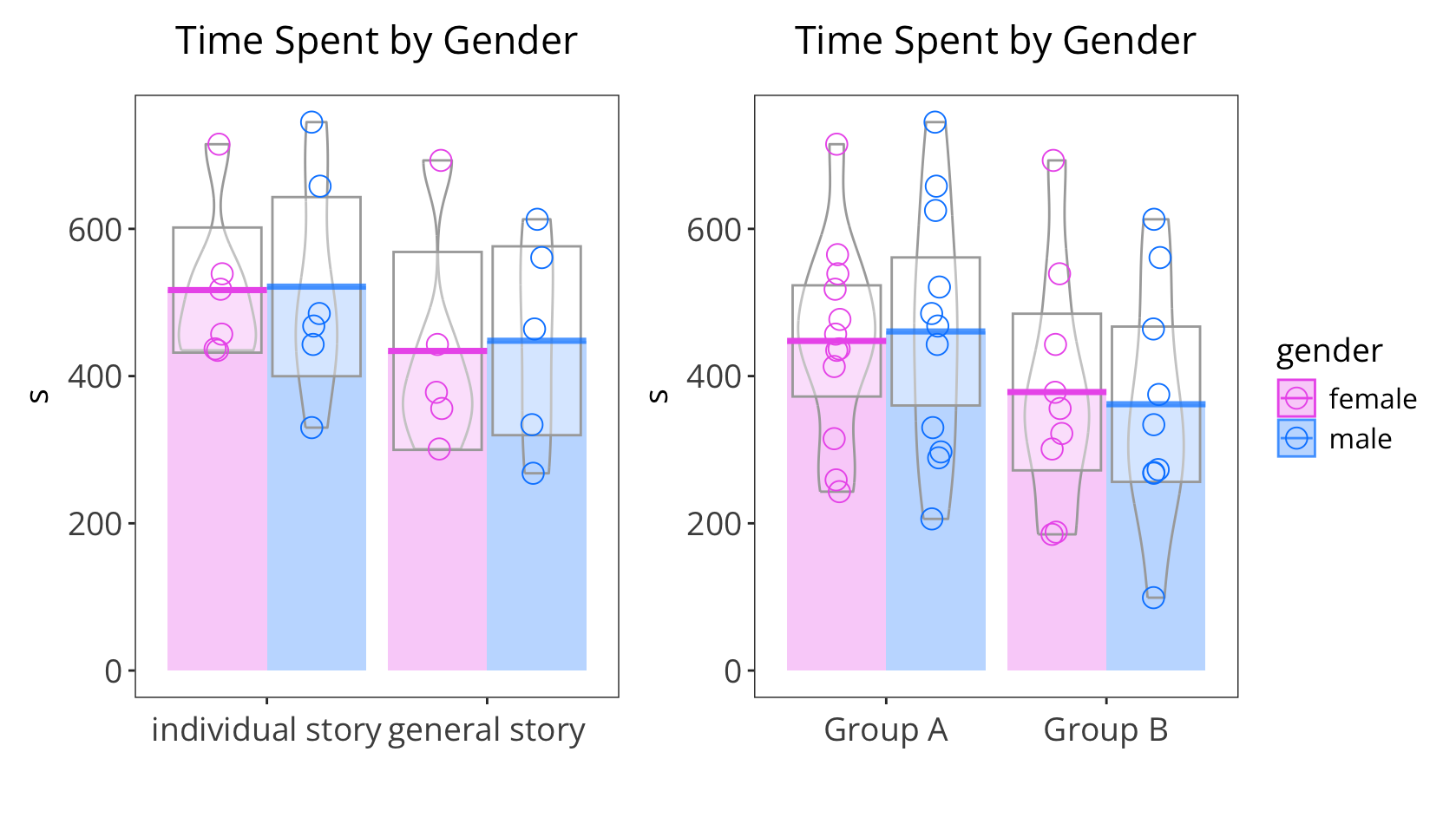}
    
  \caption{Negligible differences in time spent with respect to gender and main effects for story version and among participant groups maintained. }
  \label{fig:veri-time}
\end{figure}

\section{Effect Verification in Subgroups}
\label{app:subgroupveri}

Due to the limited sample size, demographic variables were not included in the inferential statistical analyses. Instead, a descriptive robustness check was conducted, in which the key EDA measures (Fig.~\ref{fig:veri-a} and Fig.~\ref{fig:veri-b}) and time spent (Fig.~\ref{fig:veri-time}) were visualized separately by gender. The observed main effects (stronger reactions to individual stories compared to general stories) and the priming effects between Groups A and B were consistent across the subgroups.

\end{document}